\documentclass{aa}  

\usepackage{graphicx}
\usepackage{txfonts}
\usepackage{color}
\usepackage{threeparttable}
\usepackage{tablefootnote}

\newcommand{\artemis}{ArT\'eMiS}

\newcommand{\msol}{M$_\sun$}
\newcommand{\mic}{$\,\mu$m}

\begin{document} 

   \title{Probing the structure of a massive filament: ArT\'eMiS 350 and 450~$\mu$m
   mapping of the integral-shaped filament in Orion A\thanks{This publication is based
   on data acquired with the Atacama Pathfinder Experiment (APEX), project 098.F-9304.
   APEX is a collaboration between the Max-Planck-Institut f\"ur Radioastronomie, the
   European Southern Observatory, and the Onsala Space Observatory. The temperature
   and column density maps are available in electronic form at the CDS via anonymous ftp
   to cdsarc.u-strasbg.fr (130.79.128.5) or via
   http://cdsweb.u-strasbg.fr/cgi-bin/qcat?J/A+A/
   }}

   \author{F. Schuller\inst{1,2} \and Ph. Andr\'e\inst{1} 
     \and Y. Shimajiri\inst{3,1} \and A.~Zavagno\inst{4} 
     \and N. Peretto\inst{5} \and D. Arzoumanian\inst{6,4} 
     \and T.\,Csengeri\inst{7} \and V. K\"onyves\inst{8} \and P.\,Palmeirim\inst{6}
     \and S.\,Pezzuto\inst{9}  \and A.\,Rigby\inst{5} \and H. Roussel\inst{10} 
     \and H.\,Ajeddig\inst{1} \and L.\,Dumaye\inst{1} \and P.\,Gallais\inst{1}
     \and J.\,Le Pennec\inst{1}  \and J.\,Martignac\inst{1} \and M.\,Mattern\inst{1}
     \and V.\,Rev\'eret\inst{1} \and L.\,Rodriguez\inst{1} \and M.\,Talvard\inst{1}
          }

   \institute{Laboratoire d'Astrophysique (AIM), CEA, CNRS, Universit\'e Paris-Saclay, Universit\'e Paris Diderot, Sorbonne Paris Cit\'e, 91191 Gif-sur-Yvette, France
     \and Leibniz-Institut f\"ur Astrophysik Potsdam (AIP), An der Sternwarte 16, 14482 Potsdam, Germany - \email{fschuller@aip.de}
     \and National Astronomical Observatory of Japan, Osawa 2-21-1, Mitaka, Tokyo 181-8588, Japan
    \and Universit\'e Aix Marseille, LAM \& Institut Universitaire de France, 38 rue F. Joliot-Curie, F-13388 Marseille CEDEX 13, France
    \and Cardiff University, School of Physics \& Astronomy, Queen's buildings, The parade, Cardiff CF24 3AA, UK
    \and  Instituto de Astrof\'isica e Ci{\^e}ncias do Espa\c{c}o, Universidade do Porto, CAUP, Rua das Estrelas, PT4150-762 Porto, Portugal
    \and Laboratoire d'Astrophysique de Bordeaux, Univ. Bordeaux, CNRS, B18N, all\'ee Geoffroy Saint-Hilaire, 33615 Pessac, France
    \and Jeremiah Horrocks Institute, University of Central Lancashire, Preston PR1 2HE, UK
    \and INAF-IAPS, Via Fosso del Cavaliere 100, Rome, Italy
    \and Institut d'Astrophysique de Paris, Sorbonne Universit\'e, CNRS (UMR 7095), 98 bis bd. Arago, 75014 Paris, France
 }
   \date{Received xxx; accepted xxx}

 
  \abstract
   {The Orion molecular cloud is the closest region of high-mass star formation.
   It is an ideal target for investigating the detailed structure of massive star-forming
   filaments at high resolution and the relevance of the filament paradigm 
  for the earliest stages of intermediate- to high-mass star formation.}
   {Within the Orion~A molecular cloud, the integral-shaped filament (ISF) is a prominent, 
   degree-long structure of dense gas and dust with clear signs of recent and ongoing 
   high-mass star formation.
   Our aim is to characterise the structure of this massive filament at moderately high
   angular resolution ($8\arcsec $ or $\sim \,$0.016\,pc) in order to measure the
   intrinsic width of the main filament, 
   down to scales well below 0.1~pc, which has been identified as the characteristic width of filaments.}
   {We used the ArT\'eMiS bolometer camera at APEX to map a $\sim$0.6$\times$0.2~deg$^2$
   region covering OMC-1, OMC-2, and OMC-3 at 350 and 450 \mic. We combined these data with
   {\it Herschel}-SPIRE maps to recover extended emission.
   The combined {\it Herschel}-\artemis\ maps provide details on the distribution of dense   cold material, with a high spatial dynamic range, from our $8\arcsec $ resolution up to
   the transverse angular size of the map, $\sim \,$10--15$\arcmin$. 
   By combining {\it Herschel} and \artemis\ data at 160, 250, 350, and 450\mic,
   we constructed high-resolution temperature and H$_2$ column density maps.
   We extracted radial intensity profiles from the column density map
   in several representative portions of the ISF,
   which we fitted with Gaussian and Plummer models to derive their intrinsic widths.
   We also compared the distribution of material traced by \artemis\ with that seen in the higher-density tracer N$_2$H$^+$(1--0) that was recently observed with the ALMA interferometer. }
   {All the radial profiles that we extracted show a clear deviation from a Gaussian, with evidence
   for an inner plateau that had not previously been seen clearly using {\it Herschel}-only data.
   We measure intrinsic half-power widths in the range 0.06 to 0.11 pc.
   This is significantly larger than the Gaussian widths measured for fibres seen in
   N$_2$H$^+$, which probably only traces the dense innermost regions of the
   large-scale filament.
   These half-power widths are within a factor of two of the value of $\sim \,$0.1\,pc found
   for a large sample of nearby filaments in various low-mass star-forming regions,
   which tends to indicate that the physical conditions governing the fragmentation of
   pre-stellar cores within transcritical or supercritical filaments
   are the same over a large range of masses per unit length.
   }
   {}


  \titlerunning{ArT\'eMiS mapping of the Integral-Shaped Filament in Orion A}
   \maketitle
%

\section{Introduction}

It is well established that star formation takes place in dense molecular clumps embedded
in large molecular clouds. The evolution of these dense clumps towards the onset of
gravitational collapse and the formation of solar-type stars is relatively well understood in outline, 
but several key questions remain unanswered, including how the star formation rate and efficiency
depend on the environmental conditions and whether there are notable differences in the
physical processes leading to the formation of low-mass and high-mass stars.

The {\em Herschel} Space Observatory has revealed ubiquitous  filamentary structures in the cold interstellar medium \citep[ISM; e.g.][]{ref-hgbs,Molinari2010,Arzou2011,Arzou2018,HiGAL2020}.
Many subsequent studies, including theoretical work \citep[e.g.][]{Federrath2016}, have confirmed how important filaments 
are to the evolution of interstellar matter and star formation activity, 
and they have supported a filament paradigm for star formation \citep[see the review by][]{andre2014}. 
However, most detailed (resolved) studies to date have concentrated on a few,
typically nearby clouds, where mostly low-mass and solar-type stars are forming
\cite[][among others]{Arzou2011,Hacar2013,Konyves2015}.

The Orion~A giant molecular cloud is the nearest site of active star formation
that includes high-mass stars. At a distance of only 410~pc \citep{menten2007,stutz2018b},
it is an ideal target for investigating the initial phases of high-mass star formation
in great detail. 
It is located within the Barnard's Loop, which is part of a large complex of bubbles
and filaments in a superbubble \citep{Ochs2015}.
A large filamentary structure with a length of several tens of~parsecs is seen along the north-south direction across Orion~A.
The densest part of this filament is a $\sim$1.5~deg long structure ($\sim$10~pc if in
the plane of the sky) known as the integral-shaped filament \citep[ISF;][]{bally1987}.
This structure is clearly visible in dust continuum emission \citep{Nutter2007,Shimajiri2011}
or in CO molecular lines \citep{Kong2018,Suri2019}.
The Orion Nebula Cluster (ONC) appears approximately projected on the middle of this dense filament.

On large physical scales (0.05 to 8.5~pc), \citet{Stutz2016} concluded from the
analysis of $Herschel$ data at $\sim$36$''$ angular resolution that the radial
distribution of the gas surface density near the ISF follows a power law.
At much higher angular resolution (4.5$\arcsec$), using data in the
high-density tracer N$_2$H$^+$(1--0) line obtained with the Atacama
Large Millimeter/Sub-millimter Array (ALMA), \citet{Hacar2018}
argued that the ISF contains a large number of fibres organised in a complex network.
They measured emission profiles characterised by full widths at half maximum (FWHMs) in the range 0.015--0.065~pc,
with a median value of 0.035~pc, which is significantly smaller than the typical 0.1~pc half-power 
width derived from {\it Herschel} column density maps in a number of nearby molecular complexes
\citep{Arzou2011,Arzou2018}.
Since these column density maps were derived from dust thermal emission in the sub-millimetre,
they also trace lower-density material than the N$_2$H$^+$(1--0) transition.
Therefore, the question remains open as to whether the observed difference in filament widths
can be explained by the difference in the angular resolution of the data, or if they
trace different material with different physical conditions.

With an angular resolution of 8$''$ at 350\mic, more than three times better than
{\it Herschel} at the same wavelength,
the \artemis\ camera installed at the Atacama Pathfinder Experiment (APEX)  12~m telescope \citep{ref-artemis,Andre2016} is an ideal tool for probing the cold,
dusty ISM at $<$0.1~pc scale up to distances of $\sim$3~kpc. We observed a $\sim$0.1~deg$^2$
region covering the northern end of the ISF with this instrument at 350 and 450\mic.
The \artemis\ maps, combined with {\it Herschel} data, provide a detailed view, with a large spatial-dynamic 
range, of the distribution of cold gas and dust, allowing us to draw an accurate picture
of the ISF and its substructures.
We describe the data that we use in Sect.~\ref{sec-obs}.
We then present our measurements of the radial intensity profiles along several
portions of the main filament in Sect.~\ref{sec-profile}.
We discuss our results and compare them with previous findings in Sect.~\ref{sec-discu}.
Finally, we summarise our conclusions in Sect.~\ref{sec-conclu}.


\section{\label{sec-obs}Observations and data reduction}
\subsection{{\it Herschel} data}

As part of the {\em Herschel} Gould Belt Survey (HGBS) key project \citep{ref-hgbs},
a large map of 22~deg$^2$ covering the Orion A cloud and nebula was observed at 70, 160,
250, 350, and 500\mic. Details regarding the HGBS standard data reduction and calibration
procedure can be found in \citet{Konyves2015,Konyves2019}.
In particular, the emission seen on the largest scales should be accurate since zero-level
offsets as estimated from Planck and IRAS data were applied \citep[cf.][]{Bernard2010}.

Initial results from the HGBS observations towards Orion~A were published by: \citet{Roy2013},
who investigated the physical properties of the dust in this region; \citet{Stutz2013},
who made a systematic census of the reddest protostars; and \citet{Polyc2013},
who found two different core mass functions for cores within and out of the filaments.
Here, we focus on the northern part of the Orion~A cloud, which covers a large fraction
of the ISF, including \object{OMC~1}, \object{OMC~2}, and \object{OMC~3}.

The \textit{Herschel}-SPIRE parallel-mode data are affected by saturation in the vicinity of the bright 
source \object{Orion-KL}: 55 pixels cannot be used in the original ($10\arcsec$/pixel) map at 350~\mic, nor can 14 pixels be used in the original ($14\arcsec$/pixel) map at 500~\mic.
We corrected for this by using dedicated SPIRE-only data taken towards Orion-KL 
and interpolating the values measured on neighbouring pixels. 
This leads to an extra uncertainty in the flux calibration
for this region, which covers $\sim$0.7$' \times$ 2$'$, a scale well
probed by the \artemis\ observations (see below).

\subsection{ArT\'eMiS observations}

\begin{figure*}
   \centering
   \includegraphics[width=0.99\hsize]{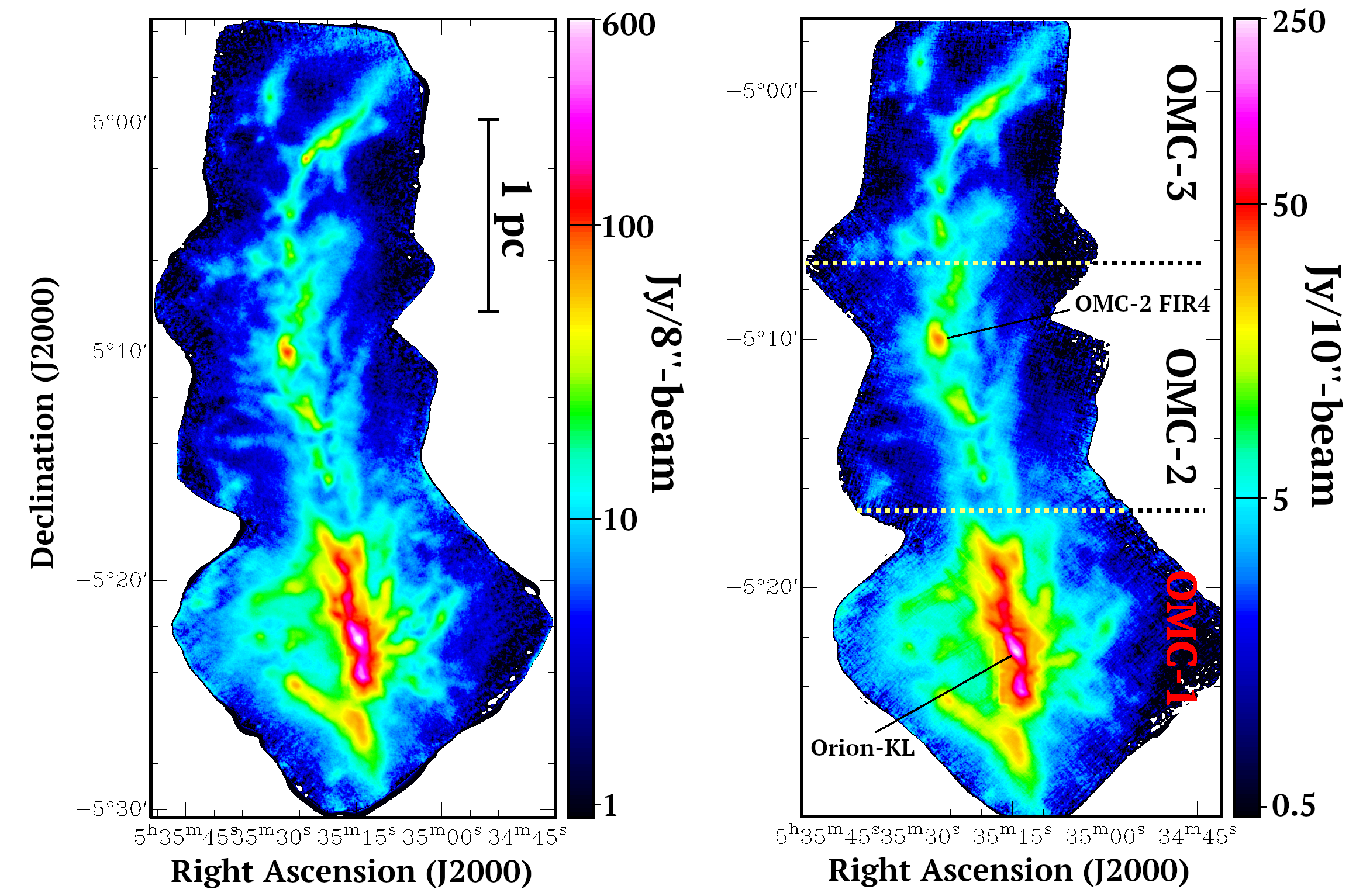}
   \caption{\artemis\ maps at 350\mic\ combined with {\it Herschel}-SPIRE data at 350\mic\ (left)
   and \artemis\ maps at 450\mic\ combined with {\it Herschel}-SPIRE data at 500\mic\ (right), shown in logarithmic colour scale. Some regions discussed in the text are indicated in the right panel.}
   \label{fig-map350}
\end{figure*}

New observations were conducted with \artemis, which is\,  installed at the APEX telescope \citep{ref-apex} in Chile.
The first maps were acquired in July 2013, during the commissioning run of the first incarnation of the camera, with only one focal plane at 350\mic.
Most observations were done with the current version of \artemis\ \citep{Talvard2018}, which covers the same $\sim$5$\times$2.5~arcmin$^2$ field of view at 350 and 450\mic, in several runs between October and December 2016.
All observations consist of large on-the-fly maps, with scanning speeds of 20$''$/s
and\, 60$''$/s and steps between lines of 5$''$ and\, 30$''$ for the data taken
in 2016 and in 2013, respectively. In total, about 16~hr were spent on-source.

The weather conditions were generally very good, with a total amount of precipitable water vapour (PWV) between 0.25 and 0.7~mm.
The zenith opacity at 350 and 450\mic\ was frequently measured with skydip observations and found to be 
between 0.8 and 1.2 at 350\mic\  and between 0.6 and 1.1 at 450\mic.
The absolute flux calibration was checked and corrected for using maps of the primary calibrator Uranus 
and several secondary calibrators (NGC~253, NGC~2071, and CRL~618).
The overall calibration uncertainty is estimated to be $\sim$30\% at 350\mic\
and $\sim$20\% at 450\mic.

\subsection{\artemis\ data reduction}

The \artemis\ data at 350 and 450\mic\ were processed with the dedicated IDL pipeline, which takes care of converting the raw data to IDL-friendly data structures, applying flux calibration, and rejecting unusable pixels. 
Special care was taken to correct for possible pointing errors between individual maps before a combined map could be computed.
Then, the data were processed with a tailored version of the \textsl{Scanamorphos} software adapted for \artemis\ data \citep{Roussel2013,Roussel2018}.
This pipeline makes use of the redundancy in the data to properly separate the various sources of noise (especially sky noise) from the real signal. This allows us to recover emission at scales that are usually filtered out by the typical data reduction of ground-based bolometer observations because such emission mimics the sky emission.

In order to recover extended emission at even larger scales, we combined the \artemis\ maps
with {\it Herschel}-SPIRE maps at similar wavelengths using the \textsl{immerge} 
task in MIRIAD~\citep{ref-miriad}.
The \textsl{immerge} algorithm combines the available data in the Fourier domain after determining an optimal calibration factor to align the flux 
scales of the input images (here from ArT\'eMiS and SPIRE) in a common annulus of the uv plane. 
Here, we adopted an annulus corresponding to the range of baselines from 
0.5~m (baseline b, sensitive to angular scales $\lambda / {\rm b} \sim 2.4\arcmin $ at $\lambda = 350\, \mu$m)  
to 3.5~m (the diameter of the {\it Herschel} telescope) to align the flux scale of the ArT\'eMiS 350\mic\ map to the flux scale of  the SPIRE 350\mic\ map.
We used the same method to combine the \artemis\ 450\mic\ data with {\it Herschel}-SPIRE maps
at 500\mic, which is the nearest wavelength observed with SPIRE.

The combined maps at 350\mic\ and 450\mic, covering about 30$\times$10~arcmin$^2$, are shown
in Fig.~\ref{fig-map350}. The effective resolution is $\sim$8$\arcsec$ and $\sim$10$\arcsec$
(FWHM) at 350 and 450\mic, respectively. 
The typical rms noise in this mosaic is between 0.3 and 0.5 Jy/8$''$-beam in regions with no strong emission at 350\mic.
The discussion in the rest of the article is based on this combined \textsl{Scanamorphos}-processed \artemis\ and SPIRE map.
We stress that the \artemis\ observations allow us not only to improve the angular
resolution compared to {\it Herschel}, but also to properly measure the peak of the emission
towards Orion-KL, where the SPIRE emission maps are saturated.

\subsection{Archival ALMA observations in N$_2$H$^+$}

\citet{Hacar2018} combined newly obtained ALMA data with data from the IRAM 30~m
telescope to build a map in N$_2$H$^+$(1--0) at 4.5$''$ resolution. 
They report the detection of 55 fibre-like structures, with typical FWHMs (as derived from Gaussian fitting)
in the range 0.02 to 0.06~pc.
We discuss the associations of these fibre-like structures with the large-scale filament seen
in sub-millimetre continuum in Sect.~\ref{sec-fiber}.

\section{\label{sec-profile}Radial profile analysis}

\begin{figure}
   \centering
   \includegraphics[width=\hsize]{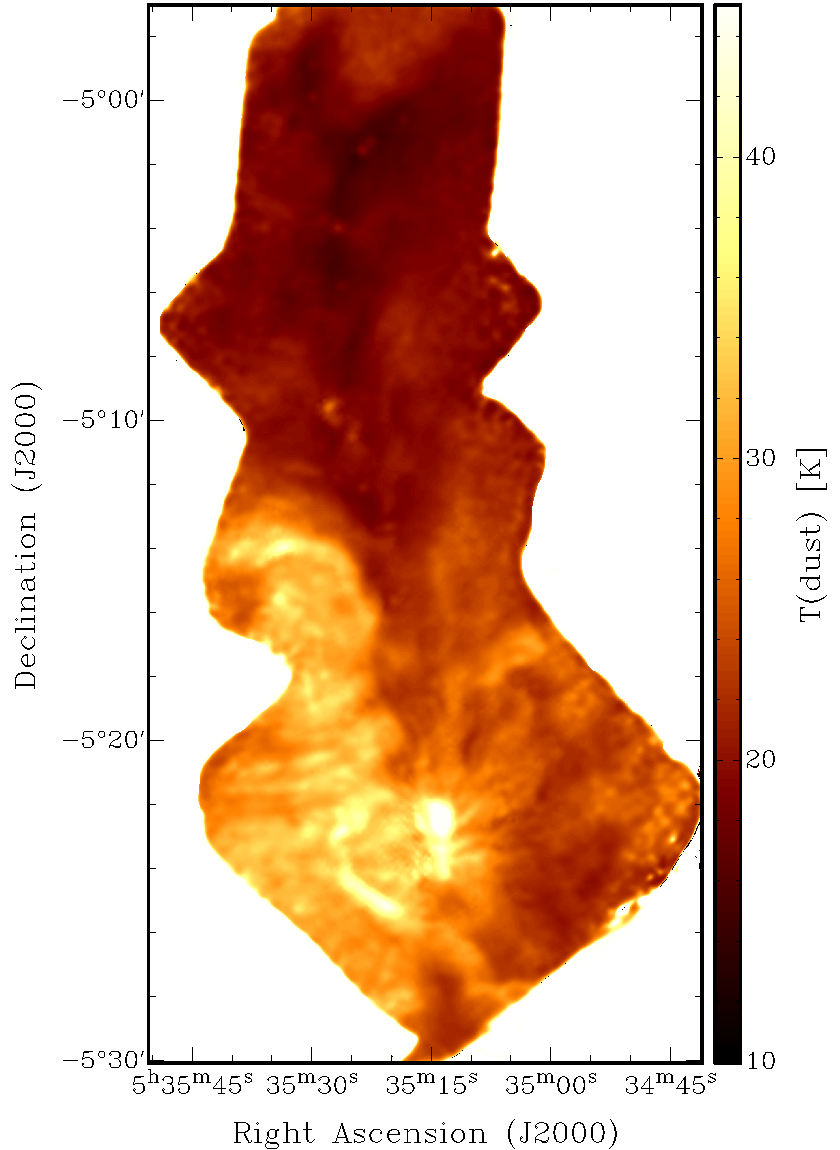}
   \caption{Dust temperature map at 18$''$ resolution, derived from the combined analysis of  \artemis\, 
   and \textit{Herschel}--PACS$+$SPIRE data.}
   \label{fig-Tdust}
\end{figure}

\begin{figure}
   \centering
   \includegraphics[width=\hsize]{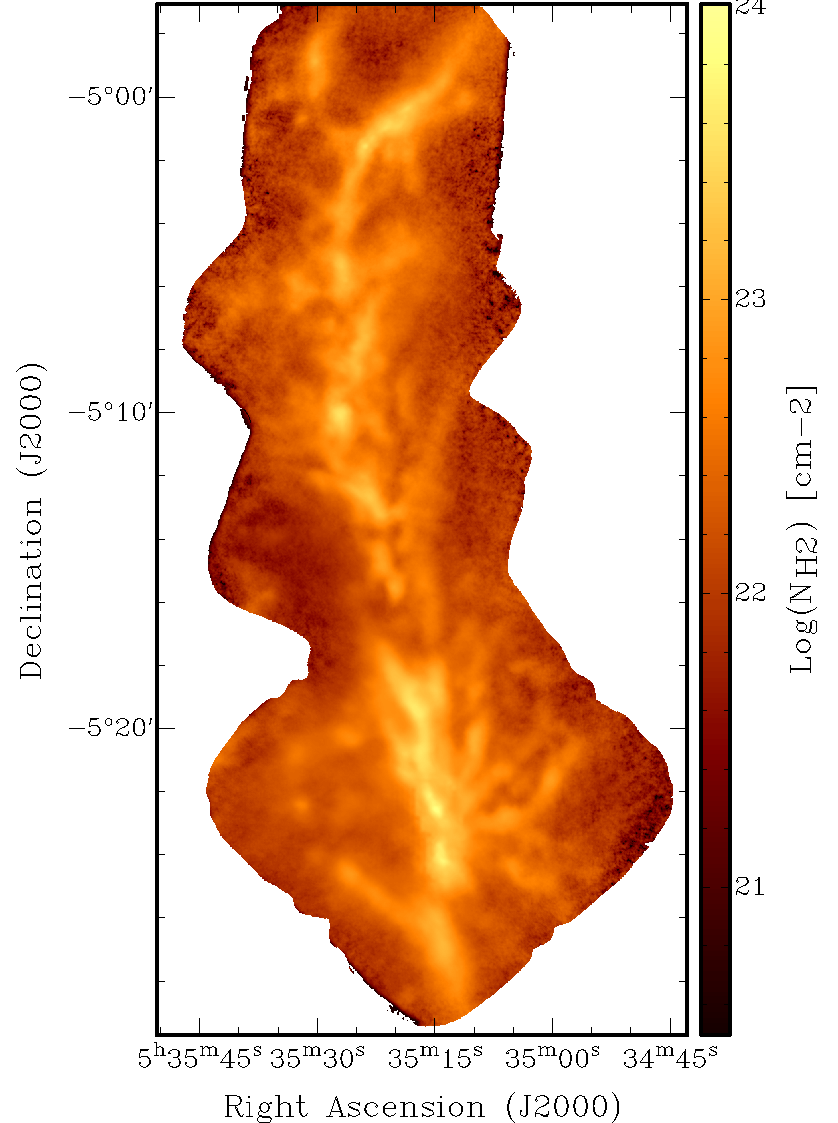}
   \caption{Column density map at $\sim$$8\arcsec $ resolution, derived from the 350\mic\ emission 
   map and the temperature map shown in Fig.~\ref{fig-Tdust},  assuming the uniform dust opacity law given by Eq.~(1) 
   (see text for details).}
   \label{fig-coldens}
\end{figure}

\begin{figure}
   \centering
   \includegraphics[width=\hsize]{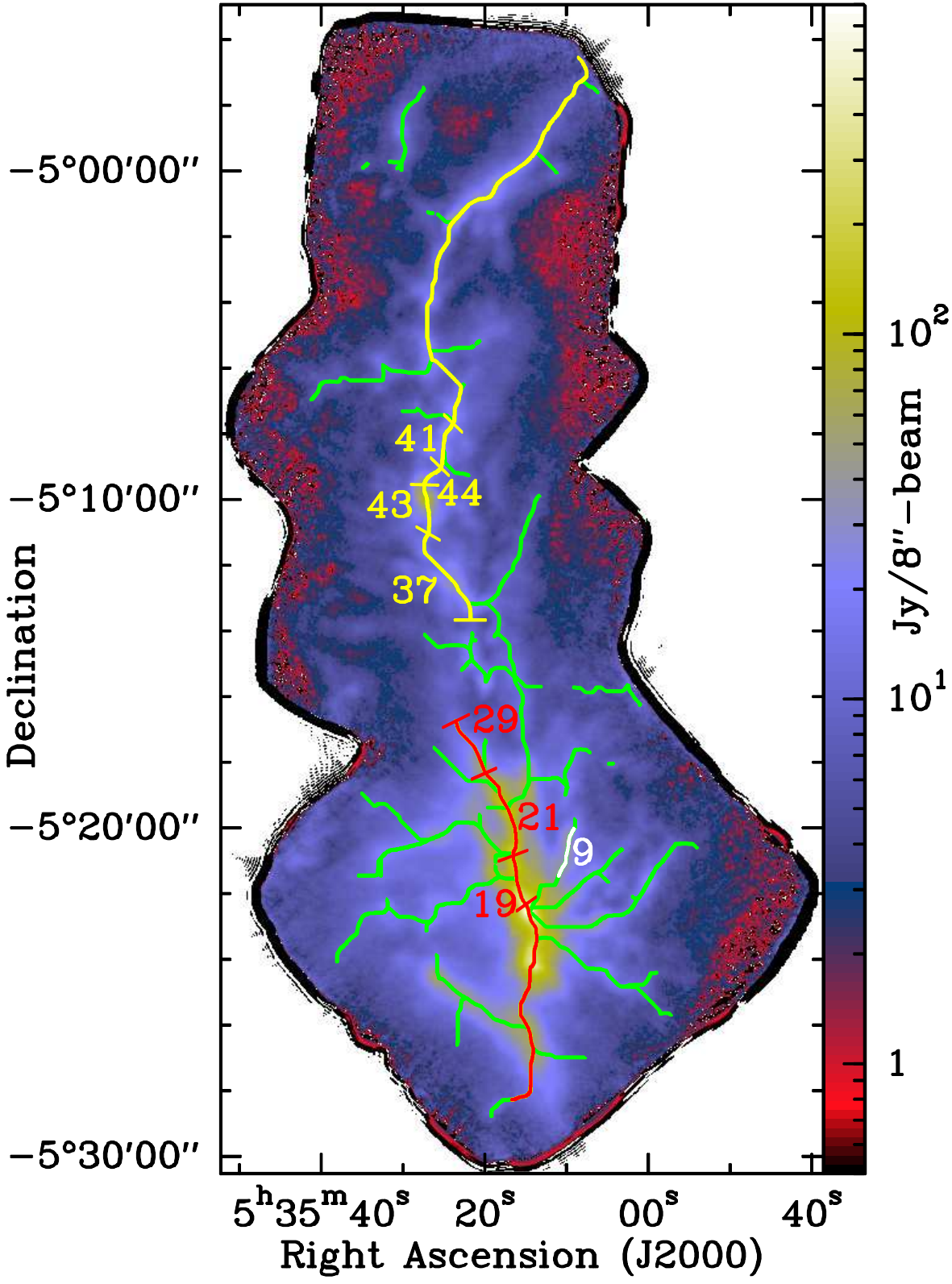}
   \caption{Skeleton map with the filament crests identified with 
   DisPerSE overlaid on the \artemis\ $+$SPIRE 350\mic\ map (logarithmic colour scale).
   The numbers mark filament segments roughly corresponding to the N$_2$H$^+$ fibres discussed by \citet{Hacar2018}, as listed in Col.~1 of Table~\ref{tab-width}. The yellow crest corresponds to the ISF portion  referred to as `ISF-OMC-2/3'
   in the text, and the red one corresponds to `ISF-OMC-1'.}
   \label{fig-skel350}
\end{figure}

\begin{figure*}
   \centering
   \includegraphics[width=\hsize]{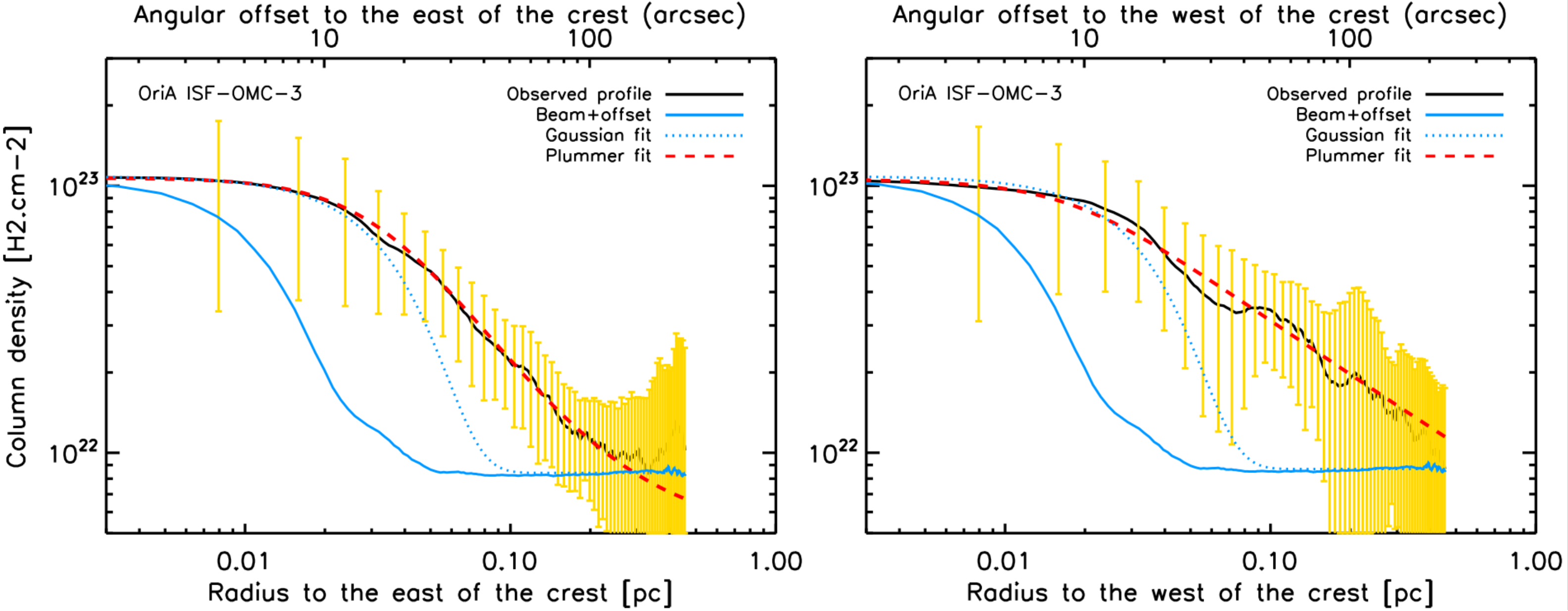}
   \caption{Median radial profiles for the northernmost part of the ISF (solid black curves), covering OMC-3, measured on the H$_2$ column density map
   perpendicular to the filament crest (see Fig.~\ref{fig-skel350}), on the eastern (left panel) and western (right panel) sides of the filament.
   One data point is shown every half beam (Nyquist sampling).
   The yellow error bars show the ($\pm 1\sigma$) dispersion of the distribution of radial profiles observed along the filament crest.
   The solid blue curves show the effective beam profile of the ArT\'eMiS 350\mic\ data shifted by a constant offset corresponding to the typical background observed nearby.
   The dotted blue curves show the best-fit Gaussian (plus constant offset) model to the inner part of the observed profile, and the dashed red curves show the best-fit Plummer model (convolved with the beam) on either side of the filament crest.
  }
   \label{fig-profile_OMC3}
\end{figure*}

\begin{figure*}
   \centering
  \includegraphics[width=\hsize]{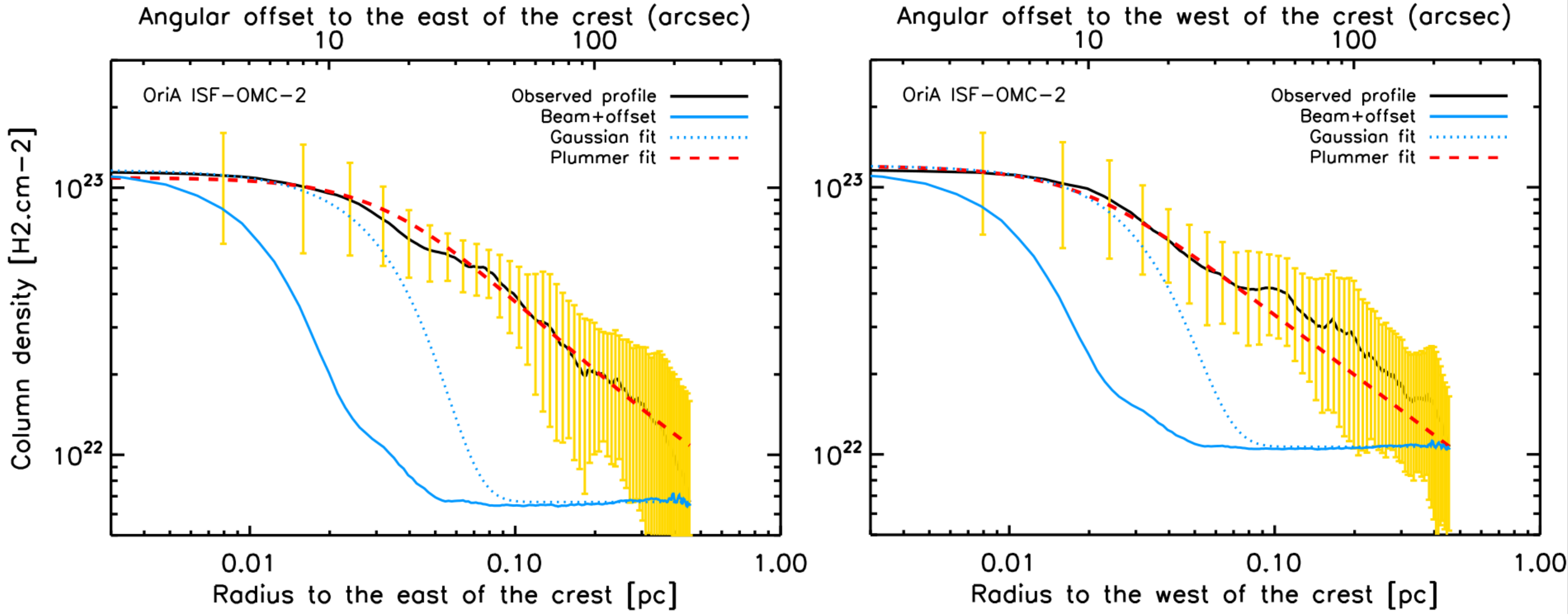}
   \caption{Same as Fig.~\ref{fig-profile_OMC3} but for the portion of
   the main filament covering the OMC-2 region.
   }
   \label{fig-profile_OMC2}
\end{figure*}

\begin{figure*}
   \centering
  \includegraphics[width=\hsize]{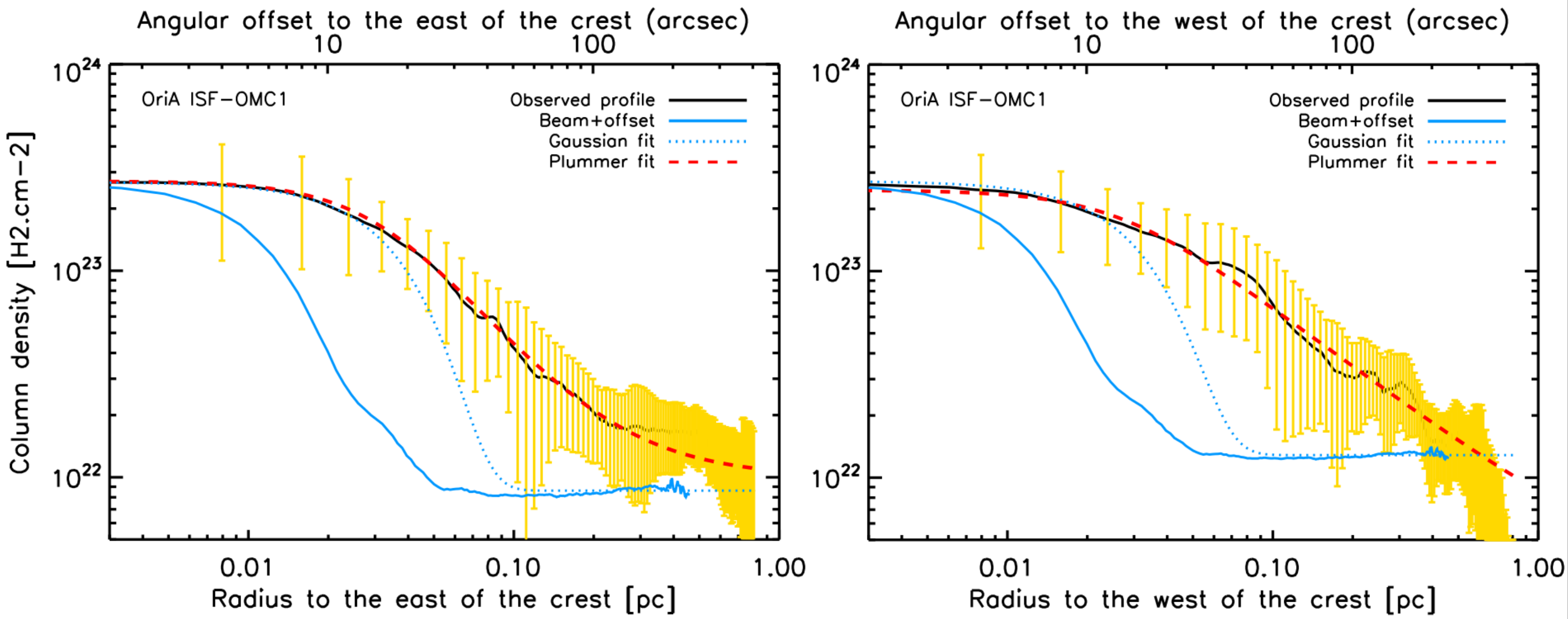}
   \caption{Same as Fig.~\ref{fig-profile_OMC3} but for the southern part of the
   main filament (including the OMC-1 region). 
   }
   \label{fig-profile_ISF_S}
\end{figure*}


\subsection{\label{sec-coldens}High-resolution temperature and column density maps}

In order to characterise the physical structures present in our maps, we first derived 
a map of H$_2$ column density at the highest possible angular resolution, using the
following method.
First, we re-projected the \textit{Herschel}-PACS, SPIRE 250~\mic, 
\artemis--SPIRE 350~\mic,\ and \artemis\ 450~\mic--SPIRE 500~\mic\ maps
to a common grid and smoothed all data to a common angular resolution of $18.2\arcsec$ 
(the lowest resolution of the above four maps, set by the SPIRE 250~\mic\ data). 
We then generated a dust temperature map at $18.2\arcsec$ resolution 
by fitting a modified blackbody of the form  $B_{\nu}(T_{\rm d}) \times \kappa_{\nu} $ 
to the four measurements between 160~\mic\ and 500~\mic\ on a pixel-by-pixel basis, 
where $B_{\nu}(T_{\rm d})$ is the Planck blackbody function at frequency $\nu$ (or wavelength $\lambda$)
for a dust temperature $T_{\rm d}$ and $\kappa_{\nu} $ (or $\kappa_{\lambda} $) is the dust opacity. 
For simplicity and easier comparison with {\it Herschel} work, our nominal assumption for 
the dust opacity law is the same as that adopted in HGBS  papers: 
\begin{equation}
\kappa_{\lambda} = 0.1 \times (\lambda/300~\mu \rm m)^{-\beta}\ \, {\rm cm}^{2}\  {\rm per\ g\ of\ gas + dust} 
\end{equation}
with an emissivity index $\beta =2$ \citep[][]{Hildebrand1983,Roy+2014}. 
The resulting dust temperature map is shown in Fig.~\ref{fig-Tdust}.
The dust temperature values range from 14~K to 25~K in the northern part  of our map, 
and from 22~K to 52~K in the southern part, with the largest values
(above 45~K) only found towards the Orion-KL \textsc{Hii} region.
The general trends seen in our map agree well with the temperature map published by
\citet{Lombardi2014} based on {\it Herschel} data only, but our data provide a better
spatial resolution and allow us to properly estimate the temperature around the
Orion-KL region, where the {\it Herschel} data are affected by saturation.

In a second step, we converted the \artemis\--SPIRE 350~\mic\ intensity map to a high-resolution
($\sim$8$''$) column density map, assuming a fixed dust temperature at each position given by 
the temperature map obtained in the first step at somewhat lower resolution. 
Assuming optically thin dust emission, the specific intensity $I_{350}$ (in MJy/sr) at $\lambda = 350$~\mic\  
was converted to H$_2$ column density as follows:
\begin{equation}
N_{\rm H_2} = I_{350}/(B_{350}[T_{\rm d}] \kappa_{350} \mu_{\rm H_2} m_{\rm H}), 
\end{equation}
where 
$\mu_{\rm H_2}=2.8$ is the mean molecular weight.  
The resulting column density map is shown in Fig.~\ref{fig-coldens} and is publicly available from the Centre de Donn\'ees astronomiques de Strasbourg (CDS). 

Uncertainties in the dust opacity law induce uncertainties in the derived temperature and column density maps. 
The temperature map does not depend on the normalisation of the dust opacity law
but is somewhat sensitive to the assumed value of the emissivity index ($\beta $), although the effect is rather small. 
For instance, changing $\beta$ from 2 to 1.5 would only increase the dust temperatures by $\sim$\, 20\% 
in the northern part and by $\sim$\, 35\% on average in the southern part of the main filament.
This would in turn decrease the derived column densities by $\sim$\,50\% and $\sim$\,70\% on average in 
the northern and southern parts, respectively.
Observationally, the spectral index map obtained between 450~\mic\ and 850~\mic\ by \citet{Johnstone1999} 
with SCUBA is consistent with $1.5 \la \beta \la 2$ in the bulk of the ISF region.
In their multi-wavelength study of the OMC--2/3 subregion, \citet{Sadavoy2016} find $\beta =$\,1.7--1.8. 
Lower, more extreme values of $\beta $ may apply locally to compact dense cores or very bright objects, such as 
OMC--3 MM6 or Orion~KL, but these values have only a limited impact on the present study, which is focused on the main filament.

On the other hand, the column density map scales directly with the adopted normalisation of the dust opacity law 
at $300~\mu \rm m$, $\kappa_{300}$. Based on a comparison of the {\it Herschel} results 
with a near-infrared extinction map of Orion~A from 2MASS, \citet{Roy2013} find evidence of a weak trend 
between dust opacity and column density, $\kappa_{300} \propto N_{H_2}^{0.28} $, in the regime 
$1 \la A_V \la 10$~mag, which they interpreted as a signature of dust grain evolution. 
Assuming this correlation between $\kappa_{300}$ and $N_{H_2}$ continues at higher $ A_V $, 
the default value assumed here for $\kappa_{300}$ should remain valid within 60\% accuracy 
for $ 3 \times 10^{21}\, {\rm cm}^{-2} \la N_{H_2} \la 10^{23}\, {\rm cm}^{-2} $
\citep[see][]{Roy+2014}. Along the spine of the integral filament where $N_{H_2} $ exceeds 
$10^{23}\, {\rm cm}^{-2} $, the actual value of $\kappa_{300}$ may be a factor of 2 to 2.5 higher 
than our default value. In turn, the map shown in Fig.~\ref{fig-coldens}  may overestimate the actual column 
density along the spine of the integral filament by a factor of $\sim$2 to 2.5. 
Significant changes in dust opacity may also occur in the Orion bar and 
the photodissociation region around the \textsc{Hii} region \citep[cf.][]{Salgado2016}, 
but these changes do not affect our analysis of the structure of the ISF away from OMC--1. 

To summarise and illustrate the influence of dust opacity uncertainties, we derived 
alternative temperature and column density maps 
from a two-parameter fitting similar to that employed for  
Figs.~\ref{fig-Tdust} and \ref{fig-coldens} but using the following alternative opacity law instead 
of the nominal HGBS opacity law: 
\begin{equation}
\kappa_{\lambda} = 0.1 \times (N_{\rm H_2}/N^0_{\rm H_2})^{0.28} \times (\lambda/300~\mu \rm m)^{-\beta(N_{\rm H_2})}\ {\rm cm}^2 {\rm g}^{-1}
,\end{equation}
where $N^0_{\rm H_2} = 1.7 \times 10^{22}\, {\rm cm}^{-2} $ and 
$\beta(N_{\rm H_2}) = \{2 \times [100-(N_{\rm H_2}/10^{21})] + 1.5 \times [(N_{\rm H_2}/10^{21})-1]\}/99 $ (i.e. 
a dust emissivity index decreasing linearly between $\beta = 2$ at $N_{\rm H_2} = 10^{21}\,  {\rm cm}^{-2}$ 
and $\beta = 1.5$ at $N_{\rm H_2} = 10^{23}\,  {\rm cm}^{-2}$). 
This alternative opacity law combines the trend found by \citet{Roy2013} 
with a plausible variation in $\beta $ with column density.  
We show the results in Appendix~A, in the form of maps of relative difference 
in temperature,  $(T_d^{\rm alt} - T_d^{\rm std})/T_d^{\rm std}$ (Fig.~\ref{fig-err-Tdust}),
and column density, $(N_{\rm H_2}^{\rm alt} - N_{\rm H_2}^{\rm std})/N_{\rm H_2}^{\rm std}$ (Fig.~\ref{fig-err-coldens}), 
between the alternative maps ($T_d^{\rm alt}$,  $N_{\rm H_2}^{\rm alt}$) 
derived using Eq.~(3) and the nominal maps ($T_d^{\rm std}$, $N_{\rm H_2}^{\rm std}$) derived using Eq.~(1). 
As expected, the alternative dust temperature map has slightly higher values 
than the nominal dust temperature map (Fig.~\ref{fig-Tdust}), by $\sim$\,10--20\% 
in the bulk of the integral filament 
and up to $\sim$\,50\% towards Orion~KL  (see Fig.~\ref{fig-err-Tdust}). 
In the OMC-1 subregion, and especially 
in the immediate vicinity of Orion~KL, the spectral energy distributions (SEDs) peak at wavelengths shorter 
than $160\, \mu$m and the dust temperature derived from measurements made longward of $160\, \mu$m is more uncertain.
Accordingly, the alternative column density map has lower values 
than the nominal column density map (Fig.~\ref{fig-coldens}) 
by up to a factor of $\sim$\,3--4 towards Orion~KL 
and only a factor of $\sim$\,2 along the bulk of the integral filament 
(see Fig.~\ref{fig-err-coldens}).

\subsection{Extraction of filamentary structures}

We applied the DisPerSE algorithm \citep{ref-disperse} 
to a version of the ArT\'eMiS$+$SPIRE 350~\mic\ image smoothed to 16$\arcsec$ resolution
to trace the crests of the most prominent filamentary structures seen in our data, including the spine of the ISF.
Apart from the main filament (i.e. the ISF), DisPerSE identified a network of fainter filamentary structures, most of which 
correspond to sub-filaments or branches directly connected to the ISF, especially in the OMC-1 region (see Fig.~\ref{fig-skel350})

In the present paper, we mostly focus our analysis on the density structure of the main filament, 
whose DisPerSE  crest is highlighted 
in red and yellow in Fig.~\ref{fig-skel350}. 
We stress that the definition of this main crest depends only weakly on the algorithm used to trace filamentary structures. 
In particular, we verified that the crests obtained with the alternative algorithms \textsl{FilFinder} \citep{Koch2015} 
and  \textsl{getsf} \citep{Menshchikov2021} are very similar to that shown in Fig.~\ref{fig-skel350} from DisPerSE (cf. 
Fig.~\ref{fig-skel-ISF} in Appendix~B). 
Indeed, filament-finding algorithms tend to yield consistent results for high-contrast filaments such as the ISF.

We constructed radial density profiles by taking perpendicular cuts at each
pixel of the $8\arcsec$ resolution column density map along each filament
crest, using a similar method to that used by  \citet{Andre2016} for \object{NGC 6334}.
The column density profiles derived for the northern (OMC-2 and OMC-3) and southern
(OMC-1) parts of the main filament (the ISF) are shown in Figs.~\ref{fig-profile_OMC3},
\ref{fig-profile_OMC2}, and \ref{fig-profile_ISF_S}.
We also investigated a smaller, relatively isolated filament (marked with the number 9 in Fig.~\ref{fig-skel350}) in
the south-western part of our map (Figs.~\ref{fig-fil15} and \ref{fig-profile_fil15}).

\subsection{\label{sec-method}Fits to the radial profiles}

To estimate the physical width of each filament, we fitted the median column density
profiles with both Gaussian and Plummer-like functions. 
We fitted the two sides of each filament separately to account for strongly varying background emission.
We also cut the main filament into three portions, corresponding to the OMC-1, OMC-2, and
OMC-3 subregions as marked in Fig.~\ref{fig-map350}, to account for the varying physical
conditions between the northern and southern part of our map.
While it is clear that other nearby features can affect individual measurements, the use of 
median profiles allows us to derive the average global properties of each portion of the main filament. 

\begin{table*}
\begin{minipage}{\linewidth}
\caption[]{Median inner widths of selected filaments, as derived from both Plummer
and Gaussian fitting to the observed radial column density profiles,
and power-law indices of the best-fit Plummer profiles.
The fibre numbers (Col.~1) refer to the fibres discussed by \citet{Hacar2018} that we tentatively associate with some segments of the filaments seen with \artemis\ (Sect.~\ref{sec-fiber}).
Columns~2 to 5 give the parameters of the best-fit models on the H$_2$ column
density map (Fig.~\ref{fig-coldens});
the values in Cols.~6 to 8 were extracted from Fig.~10 in \citet{Hacar2018} and refer to the structures seen in N$_2$H$^+$. }
  \label{tab-width}
  \begin{tabular}{l|llll|lll}
    \hline
     Fibre/& \multicolumn{4}{c}{N$_{H2}$ map (\artemis\ + {\it Herschel})} & \multicolumn{3}{c}{FWHM \citep{Hacar2018}} \\
     Filament & $D_{\rm flat}$ & $p$  &  FWHM  & $D_{\rm HP}^{\rm Plummer}$ & min & max & median \\
     & (pc) & & (pc) & (pc) & (pc) & (pc) & (pc) \\
    \hline
    \hline
    ISF-OMC-3 & & & & & & & \\
    West & 0.04$\pm$0.01 & 1.7$\pm$0.3 & 0.06$\pm$0.02 & 0.08$\pm$0.02 & & & \\
    East & 0.05$\pm$0.01  & 2.3$\pm$0.3 & 0.06$\pm$0.01 & 0.08$\pm$0.01 & & & \\
    \hline
    ISF-OMC-2 & & & & & & & \\
    West & 0.045$\pm$0.01 & 1.7$\pm$0.3 & 0.06$\pm$0.02 & 0.08$\pm$0.02 & & & \\
    East & 0.06$\pm$0.01  & 2.0$\pm$0.2 & 0.06$\pm$0.01 & 0.11$\pm$0.02 & & & \\    
    \hline
    Fibre 41 (W) & $0.045\pm0.01$ & 2.0$\pm$0.2  & 0.05$\pm$0.01 &  0.08$\pm$0.01 & 0.025 & 0.095 & 0.04 \\
    Fibre 41 (E) & $0.06\pm0.01$ & 2.2$\pm$0.3  & 0.05$\pm$0.02 &  0.09$\pm$0.02 & & & \\        
    \hline
    Fibre 44 (E) & ($0.06\pm0.02$) & (2.0$\pm$0.3)  & 0.06$\pm$0.03  & 0.10$\pm$0.03 & 0.045 & 0.065 & 0.06 \\    
    \hline
    Fibre 43 (W) & ($0.04\pm0.02$) & (2.0$\pm$0.3)  & 0.05$\pm$0.01 &  (0.07$\pm$0.02) & 0.02 & 0.06 & 0.035 \\
    Fibre 43 (E) & ($0.05\pm0.02$) & (2.3$\pm$0.3)  & (0.08$\pm$0.02) &  (0.07$\pm$0.02) & & & \\    
    \hline
    Fibre 37 (W) & ($0.07\pm0.03$) & (2.0$\pm$0.5)  & (0.07$\pm$0.02) &  (0.09$\pm$0.04) & 0.03 & 0.11 & 0.05 \\
    Fibre 37 (E) & $0.09\pm0.03$ & 2.4$\pm$0.3  & 0.14$\pm$0.03 &  0.11$\pm$0.03 & & & \\    
    \hline
    \hline
    ISF-OMC-1 & & & & & & & \\
    West & 0.05$\pm$0.01 & 2.0$\pm$0.1 & 0.055$\pm$0.01 & 0.085$\pm$0.01 & & & \\
    East & 0.05$\pm$0.02 & 2.4$\pm$0.3 & 0.06$\pm$0.01 & 0.06$\pm$0.02 & & & \\
    \hline
    Fibre 29 (W) & ($0.05\pm0.02$) & (2.3$\pm$0.5)  & 0.05$\pm$0.01 &  (0.07$\pm$0.02) & 0.025 & 0.040 & 0.035 \\ 
    Fibre 29 (E) & ($0.06\pm0.02$) & (2.5$\pm$0.5)  & 0.07$\pm$0.01 &  0.11$\pm$0.02 & & & \\
    \hline      
    Fibre 21 (W)  & $0.04\pm0.01$ & (2.0$\pm$0.3)  & 0.05$\pm$0.01 &  0.07$\pm$0.01 & 0.015 & 0.090 & 0.030 \\ 
    Fibre 21 (E)  & ($0.04\pm0.02$) & (2.0$\pm$0.5)~\tablefootmark{a}   & 0.05$\pm$0.02 &  0.06$\pm$0.01 & & & \\
    \hline
    Fibre 19 (W) & ($0.06\pm0.01$) & (2.0$\pm$0.3)~\tablefootmark{a}   & 0.07$\pm$0.02 &  0.10$\pm$0.02 & 0.005 & 0.030 & 0.020 \\
    Fibre 19 (E) & ($0.04\pm0.01$) & (2.2$\pm$0.3)~\tablefootmark{a}   & 0.06$\pm$0.01 &  0.06$\pm$0.01 & & & \\    
    \hline
    Fibre 9 (W) & ($0.04\pm0.02$) & (2.3$\pm$0.5)~\tablefootmark{a}   & 0.06$\pm$0.01 &  0.06$\pm$0.02 & 0.025 & 0.055 & 0.035 \\  
    \hline
 \end{tabular}
\end{minipage}
\begin{tablenotes}
\begin{footnotesize}
\item[] {\it Notes}: 
\item[] $-$ Unless otherwise specified, the Plummer fits on either side of each filament segment were performed over 
the whole range of radii shown in Figs.~\ref{fig-profile_OMC3} to \ref{fig-profile_ISF_S}, i.e. from 0 to $\sim\, $0.45\,pc, and  the Gaussian fits from 0 
to twice the simple half-power radius derived without any fitting (see text), i.e. typically $\sim\, $0.08--0.1\,pc.  
The western side of the OMC-2 segment was fitted with a Plummer model only out to a radius of 0.08~pc due to the presence of a shoulder in the background beyond that 
radius (see Fig.~\ref{fig-profile_OMC2}b); the eastern side of the OMC-1 segment was fitted out to a radius of 0.3~pc, and the western side of fibre 9 out to a radius of 0.1~pc.
\item[] $-$ The values shown in parentheses denote cases where the column density profile does not exhibit clear evidence of 
 a power-law wing (for example  fibre 9; see Fig.~\ref{fig-profile_fil15}), or where nearby emission strongly affects the 
 measured profile, resulting in higher uncertainties in the fitted values.
\end{footnotesize}
\end{tablenotes}
\end{table*}

Most of the filamentary structures observed here exhibit power-law emission wings, such that Gaussian functions do not reproduce their entire radial profiles well, 
as is evident in Figs.~\ref{fig-profile_OMC3}, \ref{fig-profile_OMC2}, and \ref{fig-profile_ISF_S}. 
Only the inner part of the radial profiles was thus fitted with a Gaussian function.
To do so, the width of each column density profile at half power above the background emission
was first roughly estimated using the closest point to the filament crest for which the
logarithmic derivative of the profile, $ {\rm d\, ln}\, {N_{\rm H_2}}/{\rm d\, ln}\,r $, showed a
significant switch from negative to positive (or negligible) values.
The observed profile was then fitted with a Gaussian function over a range of radii
corresponding to twice this initial width estimate.

The Plummer-like model function that was fitted to the full radial profile of 
H$_2$ column density ($N(r)$, after convolution with the approximately Gaussian $\sim 8\arcsec $ beam
of the observations) is the following:
\begin{equation}
    N_{\rm Plummer}(r) = N_0 / [1 + (r/R_{\rm flat})^2]^{\frac{p-1}{2}} + N_{\rm bg},
\end{equation}
where $R_{\rm flat}$ is the radius of the flat inner region with roughly constant emission in
the profile, $p$ is the power-law exponent of the profile at radii larger than $R_{\rm flat}$,
$N_0$ is the central column density, and $N_{\rm bg}$ is the {column density of the} 
background cloud. 
A more detailed description of the fitting procedure is given in Sect.~3 of \citet{Arzou2018}.
For a cylindrical model filament with a Plummer-like column density profile following Eq.~(4), the underlying volume density profile has a similar form:
\begin{equation}
n_{\rm Plummer}(r) = \frac{{n}_{0}}{\left[1+\left({r/R_{\rm flat}}\right)^{2}\right]^{p/2}}, 
\end{equation}
where $n_{0}$ is the central volume density of the filament. 
Moreover, $n_{0}$ is related to the projected central column density ($N_{0}$) by the simple relation 
$n_{0} = N_{0}/(A_p\, R_{\rm flat}) $, where 
$A_p = \frac{1}{\cos\,i} \times B\left(\frac{1}{2},\frac{p-1}{2}\right) $ 
is a constant factor that depends on the inclination angle ($i$) of the filament 
to the plane of sky and $ B$ is the Euler beta function \citep[cf.][]{Palmeirim2013}.

\subsection{ Results}
\label{sec-results}

%
%

We report the best-fit values of $D_{\rm flat} \equiv 2 \times R_{\rm flat}$ and $p$ from Plummer fitting, as well as
the deconvolved FWHMs from Gaussian fitting, for several segments of the ISF in Table~\ref{tab-width}.
As pointed out by several authors, the errors in $D_{\rm flat}$ and $p$ 
are significantly correlated, leading to some degeneracy in the derivation of these two parameters \citep[e.g.][]{Malinen2012, Smith2014, Arzou2018}. 
For this reason, we also provide, in the fifth column of Table~\ref{tab-width}, the half-power diameter derived from Plummer fitting:
\begin{equation}
D_{\rm HP}^{\rm Plummer} = \sqrt{2^{\frac{2}{p-1}}-1} \times  D_{\rm flat}, 
\end{equation}
defined by
\begin{equation}
N_{\rm Plummer}(D_{\rm HP}^{\rm Plummer}/2) = N_0 / 2 + N_{\rm bg}. 
\end{equation}
\noindent 
The advantage of  $D_{\rm HP}^{\rm Plummer}$ is that it is a more robust  estimator of the inner width of a Plummer-like profile 
than $D_{\rm flat}$ because its derivation is not as strongly correlated to that of $p$. 
It is worth pointing out that for high-contrast filaments  ($N_{\rm bg} << N_0$) such as the ISF,  
$D_{\rm HP}$ depends only weakly on the background column density ($N_{\rm bg}$). 
We further show in Appendix~B that $D_{\rm HP}$ is a robust quantity, affected 
by only small systematic errors ($\la 30\%$) due to uncertainties in the dust opacity 
and the presence of temperature gradients along the line of sight.

\begin{figure}
   \centering
   \includegraphics[width=\hsize]{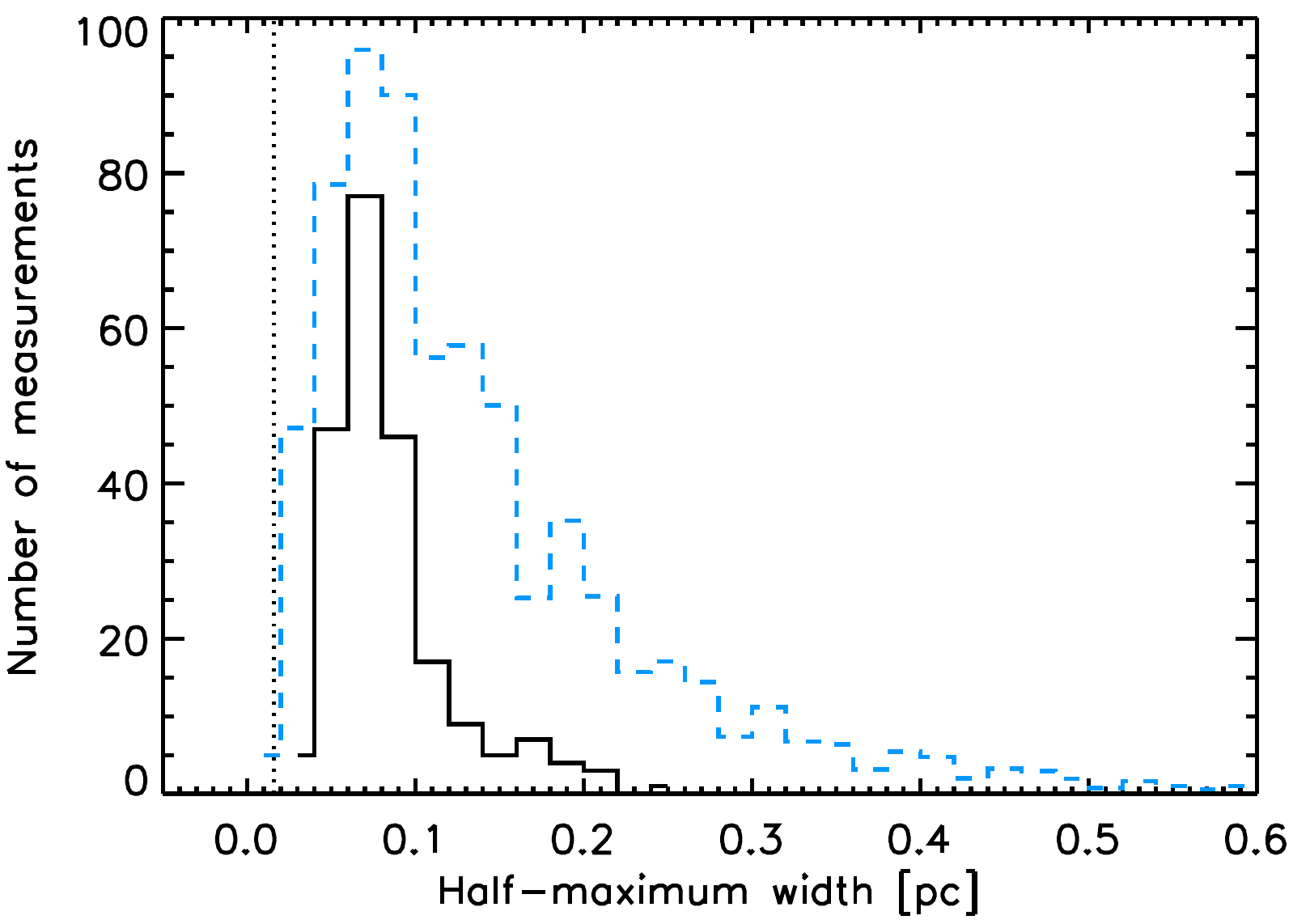}
   \caption{Histogram of 221 independent individual width measurements made 
   at beam-spaced positions along the whole (OMC-1 $+$ OMC-2 $+$ OMC-3) ISF (solid black line),
   compared to a similar histogram obtained by \citet{Arzou2018} for a large sample of 599 filaments in eight nearby clouds (dashed blue line). 
   The y-axis of the latter (blue) histogram was scaled down by a factor of 13 for easier comparison with the ISF histogram.
   The vertical dashed line marks the spatial resolution ($\sim$\,0.016\,pc) of the ArT\'eMiS 350\,$\mu$m and column density maps used in this paper 
   (Figs.~\ref{fig-map350} and \ref{fig-coldens}). 
   The median value of the OMC distribution (black histogram) is 0.075\,pc, and the interquartile range is 0.03\,pc. 
   The median value of the blue distribution is 0.1\,pc, and its interquartile range is 0.1\,pc.}
   \label{fig-width-histo}
\end{figure}

The inner widths reported in Table~\ref{tab-width} represent median values for each ISF segment, derived from a detailed fitting 
analysis of median column density profiles such as those shown in Figs.~\ref{fig-profile_OMC3}, \ref{fig-profile_OMC2}, and \ref{fig-profile_ISF_S}. 
As is evident from the yellow error bars in these figures, there are significant fluctuations in the profiles along the length of the ISF. 
To quantify the impact of these fluctuations on our filament width estimates, we display a histogram of 221 individual width measurements 
made at beam-spaced positions along the whole crest of the ISF in Fig.~\ref{fig-width-histo}. 
The individual inner widths used in Fig.~\ref{fig-width-histo} do not result from any Gaussian or Plummer fitting 
but correspond to simpler half-maximum diameters estimated as twice the radius from the filament crest 
at which the background-subtracted amplitude of the observed radial profile is half of the value on the crest 
(see `$hd$ widths' in \citealp{Arzou2018}). Such simple $hd$ width estimates are directly comparable to Gaussian FWHM or Plummer $D_{\rm HP}$ estimates, but
they are more robust and less affected by outliers. 
(When checked and reliable, Gaussian and Plummer fits are nevertheless preferable 
since they provide a better characterisation of the filament profiles and additional information such as line mass 
and density estimates.)
Figure~\ref{fig-width-histo} also compares the distribution of individual widths obtained here for the ISF 
(black histogram)
with the distribution of widths found by \citealp{Arzou2018}  (blue histogram). 
Although the latter is somewhat broader than the former, which is not surprising 
since it is based on a much broader sample of 599 filaments in eight nearby clouds at different distances, 
it can be seen that the two distributions have very similar overall shapes and peak at the same value, just below 0.1\,pc.
 
Overall, we measure inner $D_{\rm HP}^{\rm Plummer}$ widths in the range of 0.06 to 0.11~pc, 
consistent within better than a factor of two 
with the 
half-power widths reported by \citet{Arzou2018} for other Gould Belt filaments and by \citet{Andre2016} for
the NGC~6334 filament ($D_{\rm HP}^{\rm Plummer} = 0.16\pm0.05$~pc at 1.7 kpc, or $0.12\pm0.03$~pc adopting a revised distance 
of 1.3~kpc to NGC~6334; cf. \citealp{Chibueze2014}).
Particularly remarkable is the similarity in width with the Taurus B211/B213 filament ($D_{\rm HP}^{\rm Plummer} = 0.11\pm0.02$~pc -- \citealp{Palmeirim2013}),
which has an order of magnitude lower column density than the ISF.
The errors given in Table~\ref{tab-width} are only statistical fitting uncertainties. 
To assess the importance of systematic errors due to uncertainties in the dust properties, we also 
performed a similar radial profile analysis on the alternative column density map derived using Eq.~(3) instead of Eq.~(1) for the dust opacity law. 
The results are provided in Appendix~\ref{sec-app-alt} (Table~\ref{tab-alt-width}).  
Generally speaking, the measured radial profiles tend to be slightly shallower, with slightly lower $p$ index values (by $\delta p \sim 0.2)$, 
in the alternative column density map.
The inner widths derived from the alternative column density map range from 0.07 to 0.16~pc
and are typically $\sim$\,30\% higher than those found with the default column density map. 
They remain consistent with the filament widths found by \citet{Arzou2018}.
Overall, uncertainties in the dust opacity law have a stronger influence on the absolute level 
of the column density profiles than on the profile shapes and the derived filament widths. 

We also verified that our results on the widths do not depend on the choice of the algorithm employed 
to trace the crest of the ISF (see Appendix~B). In particular, a Kolmogorov-Smirnov test reveals that the three distributions 
of individual widths measured along the ISF using the DisPerSE, \textsl{FilFinder}, and \textsl{getsf} crests, respectively, 
are statistically indistinguishable (see Fig.~\ref{fig-width-getsf-filfinder}).

The inner widths that we measure are almost always larger than the 0.02--0.06~pc range of values derived by \citet{Hacar2018} 
from combined ALMA$+$IRAM~30m observations in the high-density tracer N$_2$H$^+$(1--0).
Because the N$_2$H$^+$(1--0) line is effectively tracing only dense gas with $ n_{H_2}  \ga 10^4 \, {\rm cm}^{-3} $ \citep[cf.][]{Shirley2015}, 
it may not properly sample the outer parts of the underlying radial density profiles, therefore biasing the results towards lower width values. 

The $p$ exponents of the best-fit Plummer profiles for the various portions of the ISF are 
consistent with the typical values of $p = 2.2\pm 0.3$ found by \citet{Arzou2018} 
for a large sample of nearby {\it Herschel} filaments in the Gould Belt. 
For the northern part of the ISF, they are also roughly consistent with the power-law exponent
$p = 1.5\pm 0.5$ reported by \citet{Mattern2018} for another wide sample 
of almost 200 larger-scale, velocity-coherent filaments in the inner Galaxy from
the ATLASGAL 870\mic\ and SEDIGISM $^{13}$CO(2--1) surveys\citep{Schuller2009,Schuller2017}.
These relatively shallow exponents indicate that
the main filament features pronounced non-Gaussian wings of lower-density material 
that extend well beyond 0.1~pc from the ridge, as evident in Figs.~\ref{fig-profile_OMC3} to \ref{fig-profile_ISF_S}.
The profile of ISF-OMC-1 is strongly affected by the Orion-KL star-forming
region, resulting in a somewhat steeper exponent.

The mean central H$_2$ densities inferred from our Plummer fits range from $\sim 4\times 10^5\, {\rm cm}^{-3}$ for  OMC-3 and OMC-2 
to $\sim 10^6\, {\rm cm}^{-3}$ for OMC-1, assuming the ISF is in the plane of the sky (i.e. $i =0$). 
The mean central density we infer for the OMC-1 portion of the ISF is a factor of three lower 
than the nominal volume density of $3 \times 10^6\, {\rm cm}^{-3}$ found by \citet{Teng2020} 
for N$_2$H$^+$  filamentary structures in OMC-1 (outside of cores) 
through non-LTE modelling of the N$_2$H$^+$ (3--2) and (1--0) lines. 
Our dust continuum estimate of the volume density is, however, affected by a rather large 
uncertainty of a factor of $\sim$2--3 or more in the OMC-1 subregion, which is
due to particularly uncertain dust opacity and temperature gradient effects in 
this unusually dense and strongly irradiated area 
(see e.g. Figs.~\ref{fig-err-Tdust}~and~\ref{fig-err-coldens} in Appendix~A). 
The N$_2$H$^+$ density estimates of  \citet{Teng2020} are also uncertain 
by a factor of $\sim$3 (see their Table~3). 
Given these uncertainties, the two sets of density estimates are broadly consistent. 

\begin{figure}
   \centering
   \includegraphics[width=1.0\hsize]{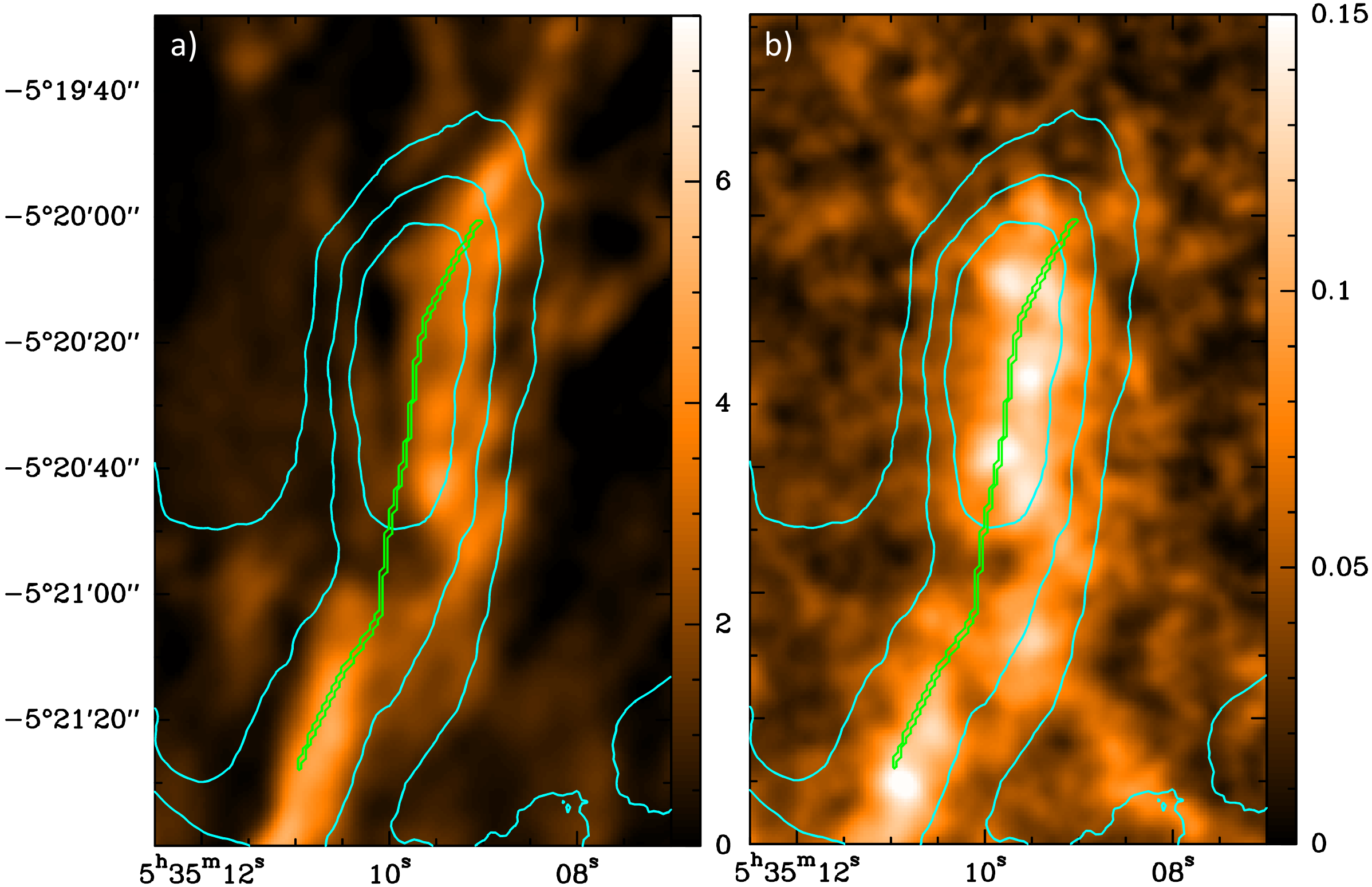}
   \caption{Comparison of our ArT\'eMiS column density map (cyan contours)
with {\bf a)} the ALMA$+$30m N$_2$H$^+$ moment-0 map from \citet{Hacar2018}
and {\bf b)} the VLA$+$GBT NH$_3$ moment-0 map from \citet{Monsch2018}
for the filament we associate with fibre 9. The ArT\'eMiS column density contours 
are 5, 7, and $9 \times10^{22}\, {\rm cm}^{-2}$. 
The green line shows the filament's crest as derived with DisPerSE.}
   \label{fig-fil15}
\end{figure}

\begin{figure}
   \centering
   \includegraphics[width=\hsize]{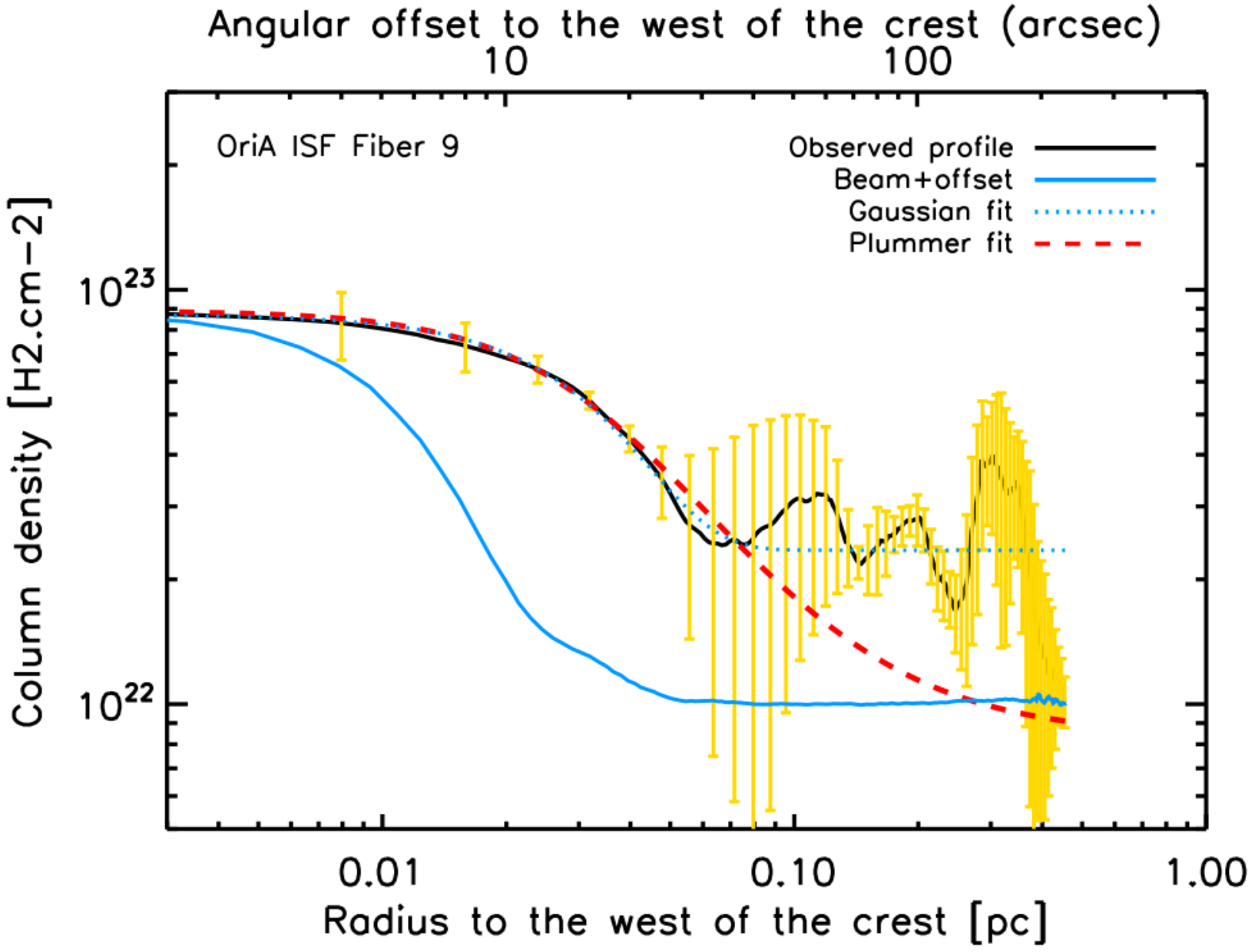}
   \caption{Same as Fig.~\ref{fig-profile_OMC3} but for the south-western side of the filament extracted with DisPerSE that coincides with fibre 9 of \citet{Hacar2018}.
}
   \label{fig-profile_fil15}
\end{figure}

By integrating the radial profiles measured in the column density map of
Fig.~\ref{fig-coldens} over radii up to $D_{\rm HP}$ (i.e. twice the half-power
radius), we also derived robust estimates of 
the mean mass per unit length of the inner portion of each ISF segment, excluding low-density material at large radii.
The HGBS results suggest that most stars form within this inner portion 
of molecular filaments \citep[e.g.][]{Konyves2015,Konyves2019}. 
We obtain average values of $\sim 200\, M_\odot $/pc for OMC-3, 
$\sim 250\, M_\odot $/pc for OMC-2, 
$\sim\,$500$\, M_\odot $/pc for the OMC-1 portion, 
and $\sim\,$300$\, M_\odot $/pc for the whole northern ISF (i.e. OMC-1$+$OMC-2$+$OMC-3).
For the filament  shown in Fig.~\ref{fig-fil15}, which can be associated with
fibre 9 of \citet{Hacar2018}, we estimate a mean line mass of $\sim 100\, M_\odot $/pc (see the column density profile in Fig.~\ref{fig-profile_fil15}).
These line mass estimates were validated via comparison with the low-resolution
($5\arcmin$) dust optical depth map of \citet{Planck_mamd2014}.
They are more than one order of magnitude higher than the line masses of 
the `fertile fibres'  observed in the nearby Taurus cloud \citep{Hacar2013} and 
of the many transcritical filaments that dominate the filament line mass function
(FLMF) in  the \citet{Arzou2018} HGBS sample (see Fig.~1 of \citealp{Andre2019}), 
but they are consistent with the bulk of confirmed Galactic filaments discussed by \citet{Mattern2018}.
Our measurements are also in agreement with the results of \citet{HiGAL2020},
who find a mean value of $250\, M_\odot $/pc for a sample of 18,000 Galactic
filaments with reliable distance estimates.
It is not surprising to find such high line mass values in the Orion ISF since it
is known to be undergoing fragmentation and ongoing embedded star formation
\citep[e.g.][]{Takahashi2013,Teixeira2016}.

%
%

\section{\label{sec-discu}Discussion}

\subsection{Density structure}

The large-scale distribution of gas and dust in the northern part of Orion~A is dominated by the ISF.
Based on a column density map at 36$''$ resolution built from HGBS data, \citet{Stutz2016} find that the radial distribution 
of gas near the ISF is well represented by a power law.
They reported a line mass profile ($\lambda (w)$, in \msol/pc) as a function of enclosing radius ($w$) that follows the following function well:
\begin{equation}
    \lambda(w) = K \left( \frac{w}{\rm{pc}} \right)^\gamma,
\end{equation}
where $K = 385\, M_\odot $/pc and $\gamma$=3/8. 
This value of $K $ was derived assuming the dust opacity law OH5 from \citet{Ossenkopf94}, 
which roughly corresponds to the same dust opacity parameterisation as adopted in
Sect.~\ref{sec-profile} but with $\beta = 1.8$ instead of $\beta = 2$. 
More precisely, the OH5 dust opacity at ArT\' eMiS wavelengths is 
a factor of $\sim$30\% higher than the HGBS opacity law of Eq.~(1). 
Equation~(8) also corresponds to a line mass profile 
averaged over the full 1$^\circ$ length of the ISF, including 
a $0.4^\circ$ long portion south of the field mapped with ArT\'eMiS, 
whose average column density is a factor of two or 
more lower than the northern part of the ISF discussed here,
according to {\it Herschel} and {\it Planck} data. 
Considering these differences, the average line mass of 
$\sim\,$300$\, M_\odot $/pc 
that we derived in Sect.~\ref{sec-results} 
for the northern part of the ISF, 
integrated over a transverse enclosing 
radius of  $D_{\rm HP} \sim $\,0.07--0.1\,pc, 
is consistent with the value of $K$ reported 
by \citet{Stutz2016} for a radius of $w=1$\,pc.

\begin{figure}
   \centering
   \includegraphics[width=\hsize]{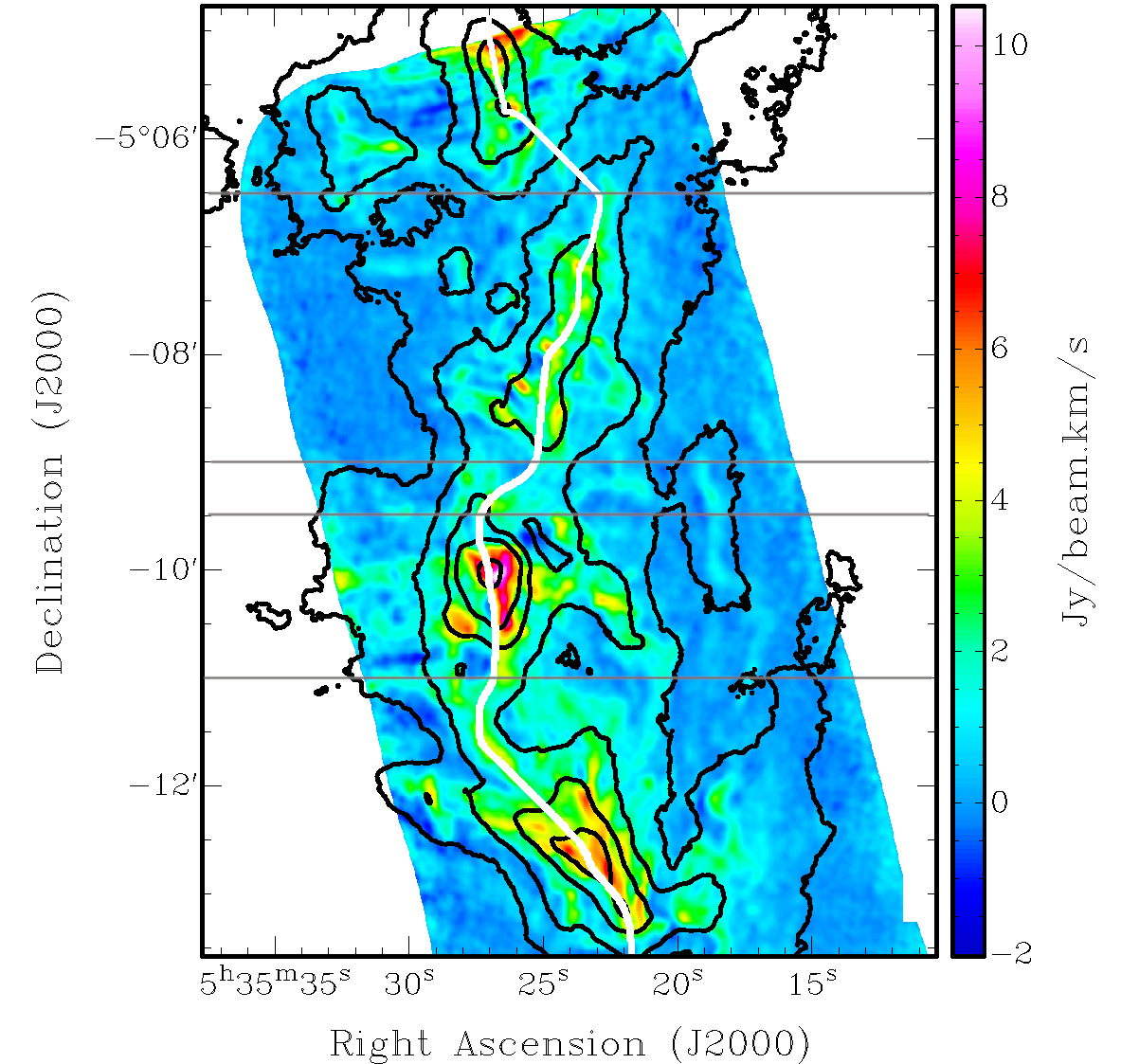}
   \caption{ALMA N$_2$H$^+$ moment-0 map \citep[from][]{Hacar2018} overlaid with
   our H$_2$ column density map (black contours) for the ISF-OMC-2 region.
   The thick white line shows the crest of the filament extracted with DisPerSE.
   The thin horizontal lines indicate the limits of the various segments discussed
   in Sect.~\ref{sec-Fibre-isf-north}.}
   \label{fig-isf-n2hp}
\end{figure}

The column density map used by \citet{Stutz2016} is sensitive to scales between
0.05 pc and 8.5 pc. Based on our higher-resolution data, we find an inner plateau with
a total half-power width of order $\sim\,$0.06~pc to 0.11~pc (see Table~\ref{tab-width}).
Therefore, while our results are consistent with the analysis of \citet{Stutz2016},
they allow us to resolve an inner break in the radial density and line mass profiles
of the ISF that could not be seen with {\it Herschel} data only. 
From their analysis at 36$''$ resolution, \citet{Stutz2016} did not find any
significant variation of the column density profile along the extent of the ISF, 
but this is not surprising since they did not resolve the inner plateau 
in the density profile.

\subsection{\label{sec-fiber}Comparison with high-density fibres}

Although it is nearly impossible to associate each N$_2$H$^+$ fibre reported by \citet{Hacar2018}
with the ArT\'eMiS/SPIRE filaments that we extracted with DisPerSE,
we will now discuss a few examples.

\subsubsection{\label{sec-Fibre-isf-north}ISF-OMC-2}
First, we focused on the northern part of the ISF, where the main filament shows a
relatively clear structure as compared to the highly complex
network of filaments around the extreme region Orion-KL.
The map published by \citet{Hacar2018} covers only
$-05\degr 25' \leq \delta \leq -05\degr 05'$; therefore, we considered only the
southern part of what we call ISF-OMC-2/3, which roughly corresponds to OMC-2.
We analysed the radial column density profiles of different
segments of the ISF, which correspond to different ranges in declination, as shown in
Fig.~\ref{fig-isf-n2hp}. For each segment, we give the parameters of the best-fit
Plummer and Gaussian models in Table~\ref{tab-width}, together with the 
range of FWHMs measured in N$_2$H$^+$ by \citet{Hacar2018} for each fibre that we tentatively associate with the \artemis\ filaments (see also Fig.~\ref{fig-skel350}
for the numbering of the fibres).

The segment from  $\delta = -05\degr 11'$ to $\delta = -05\degr 09'30''$, which
roughly corresponds to fibre 43 from  \citet{Hacar2018}, contains the bright
infrared source OMC-2~FIR~4. The values of the inner width that we measure for this segment
($D_{\rm HP}^{\rm Plummer} = 0.07 \pm 0.025 $\,pc) 
are in the lowest boundary of our sample. However, we caution the reader that this
measurement is strongly affected by the presence of the centrally condensed
protostellar source OMC-2~FIR~4 within this short segment \citep{Shimajiri2008,Lopez2013}.
Assessing the intrinsic width of the underlying parent filament at this location is more uncertain.
Nevertheless, such peculiar sources have very little effect on our estimates
of the global properties of the ISF since we evaluated them from median radial profiles.

\subsubsection{ISF-OMC-1}

In the southern part of our ArT\'eMiS map, we also split the ISF into several
segments, but only north of the {\sc Hii} region Orion-KL. We tentatively associated
these segments with fibres seen in N$_2$H$^+$ (see numbering in Fig.~\ref{fig-skel350}). The inner widths as measured from Gaussian and Plummer
model fitting are also given in Table~\ref{tab-width}, together with the FWHMs measured in the N$_2$H$^+$ data.
Here also we derive inner widths significantly smaller than 0.1~pc for some
portions of the main filament (e.g.\ the segment associated with fibre 21), where
a dense protostellar core is embedded and locally affects our measurements.

\subsubsection{Comparison between \artemis\ and N$_2$H$^+$}

\begin{figure}
   \centering
   \includegraphics[width=0.99\hsize,trim= 0.5cm 0cm 2cm 0cm, clip]{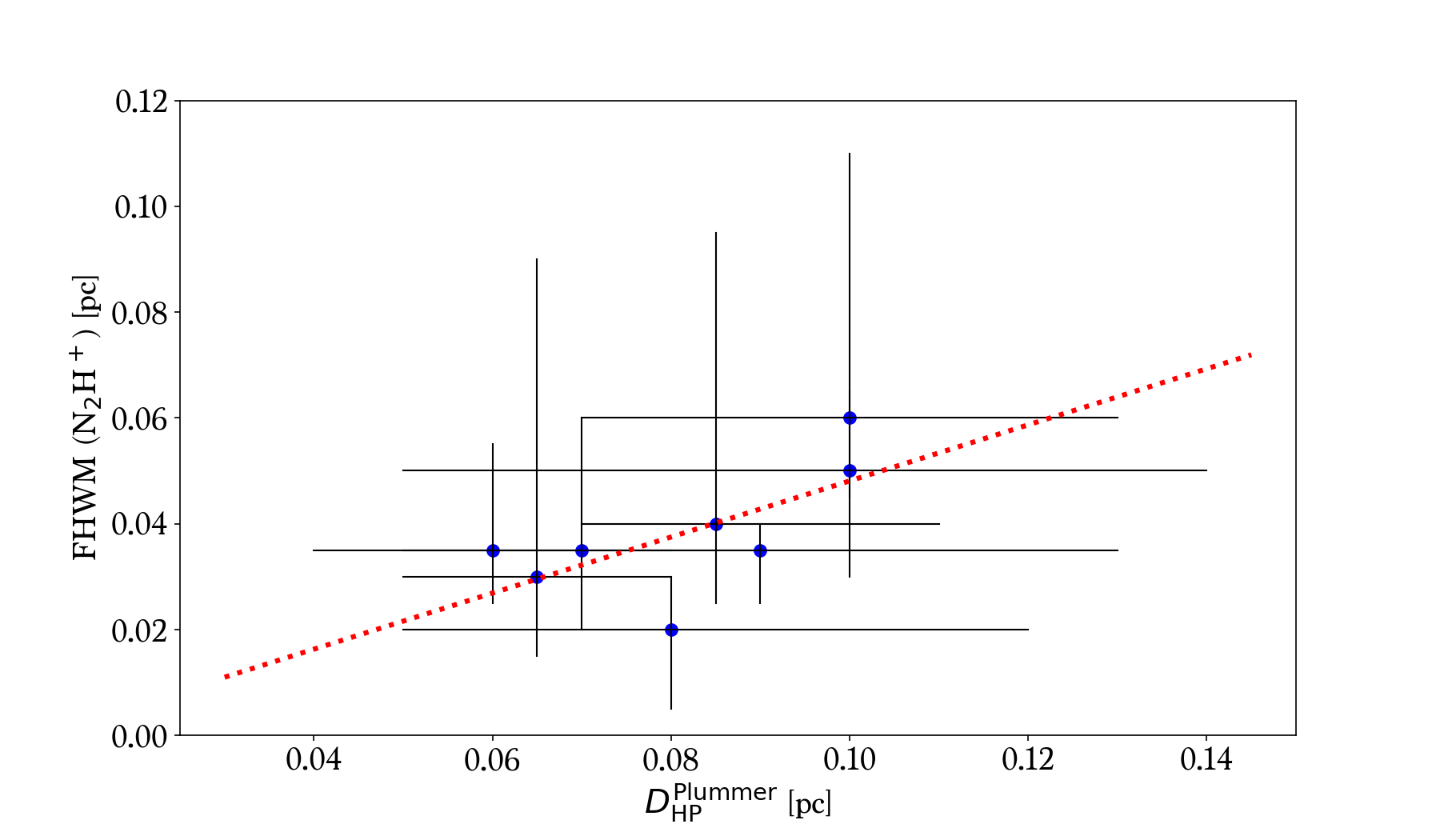}
   \caption{Comparison between the inner widths ($D_{\rm HP}^{\rm Plummer}$) measured on the high-resolution column density map (Fig.~\ref{fig-coldens})
   and the FWHM fitted on the associated fibres seen in N$_2$H$^+$.
   Each point is located at the average of the $D_{\rm HP}^{\rm Plummer}$ values
   measured eastwards and westwards of the crest vs. the median value of the fibre's
   distribution of FWHMs. The error bars indicate the full range of values on
   both axes.
   The dotted red line shows the position of a linear fit to the data.}
   \label{fig-compare-widths}
\end{figure}

We show in Fig.~\ref{fig-compare-widths} the comparison between the 
$D_{\rm HP}^{\rm Plummer}$ diameters measured on our column density map
and the FWHM of the N$_2$H$^+$ fibres that we associated with the different
segments of the ISF, as listed in Table~\ref{tab-width}.
This figure shows that there is a trend of increasing FWHM for the N$_2$H$^+$ fibres
with increasing $D_{\rm HP}^{\rm Plummer}$: A linear fit to the data results in
a slope of 0.53 ($\pm$0.24) and an intercept of -0.005 ($\pm$0.020);
the Pearson correlation coefficient is equal to 0.66.
Although the correlation is only marginal, we interpret this as an indication that 
N$_2$H$^+$ emission generally traces the densest portion along the central axis of the main filament. 

Slight but significant deviations between the peaks in N$_2$H$^+$ emission and the filament crests 
traced by our sub-millimetre dust continuum data are visible in some cases, such as the filament 
associated with fibre 9 of \citet[][see our Fig.~\ref{fig-fil15}a]{Hacar2018}. 
In that particular case, it is noteworthy that there is better agreement between the 
\artemis-SPIRE filament and the filamentary structure detected in NH$_3$  by \citet{Monsch2018} at $6\arcsec $ resolution (see \ Fig.~\ref{fig-fil15}b).
At least in projection, fibre 9 lies between two large-scale H$_2$ `fingers' 
associated with the explosive CO outflow emanating 
from the immediate vicinity  of Orion~KL \citep[cf.][]{Zapata2009,Bally2017}. 
The southern end of fibre 9 also overlaps with the area where high-velocity CO(6-5) emission 
was detected in the APEX observations of \citet{Peng2012}. 
As CO is known to be a destroyer of N$_2$H$^+$ \citep[e.g.][]{Aikawa2001}, it is possible 
that energetic finger-like outflow features may significantly reduce the abundance 
of  N$_2$H$^+$ in the lateral edges of the dense filamentary structure. 
Strong anisotropic irradiation of fibre 9 by the Orion nebula 
(see e.g. the dust temperature map shown in Fig.~\ref{fig-Tdust}) may also play a role 
via the desorption of CO molecules from grain mantles. 
In any event, a comparison of the N$_2$H$^+$ map of  \citet{Hacar2018} with the 
NH$_3$ map of \citet{Monsch2018} clearly indicates that the abundance of N$_2$H$^+$ 
is far from uniform in the OMC-1 region. The most extreme difference between the two 
maps is that no N$_2$H$^+$ emission is detected in the immediate vicinity of Orion~KL 
\citep[see also][]{Teng2020}, presumably due to the destruction of N$_2$H$^+$ by CO, 
while strong NH$_3$~(1,1) and NH$_3$~(2,2) emission is seen in the \citet{Monsch2018} data 
and a prominent column density peak is observed in our data 
(cf. Figs.~\ref{fig-map350} and~\ref{fig-coldens}). 



\section{\label{sec-conclu}Conclusions}

We have presented 350 and 450\mic\ observations with \artemis\ of the ISF in Orion, the nearest site of active high- and intermediate-mass star formation.
With angular resolutions of 8$''$ and 10$''$, the \artemis\ data, combined with {\it Herschel}-SPIRE maps to recover large-scale emission, probe scales from 0.015~pc to a few pc.
By combining the new \artemis\ data with {\it Herschel} data at shorter wavelengths,
we were able to build, for the first time, high-resolution temperature and column density maps
covering a large region around OMC-1 that are not affected by saturation; these maps
are publicly available from the CDS\footnote{Link to A\&A online or to the CDS.}.

We extracted the radial profiles and intrinsic widths of several segments of the
Orion ISF from the high-resolution column density map.
Our conclusions can be summarised as follows.
\begin{itemize}
  \item We resolve an inner plateau with a typical half-power width in the range
  0.06 to 0.11~pc along the northern and southern parts of the main filament.
  This plateau could not be seen in the {\it Herschel}-SPIRE data because of their 
  limited spatial resolution of $\sim$0.06~pc.
  \item These values of the inner width are consistent within better than a factor of two 
  with the values measured in several nearby molecular clouds of the Gould Belt 
  \citep{Arzou2011,Arzou2018,Palmeirim2013} and in the massive star-forming complex NGC~6334
  \citep{Andre2016}.
  \item The mean line masses that we derive from our data (between $\sim$100 and $\sim$ 500~$M_\odot $/pc) are in the extreme upper range of those  found for {\it Herschel} filaments in the Gould Belt \citep{Arzou2018, Andre2019}, as well as in the upper range of those measured
  for more than 18,000 filaments found in the Hi-GAL survey by \citet{HiGAL2020}.
  Our values are comparable to the line masses measured for larger-scale filaments
 throughout the Galaxy in the APEX/SEDIGISM survey \citep{Mattern2018}; 
 this is consistent with the picture in which filament fragmentation and star formation occurs 
 within the densest transcritical or supercritical portions of long, roughly $\sim \,$0.1\, pc wide filaments.
\end{itemize}
 
The present study also highlights the importance of combining data that cover a wide range of spatial scales and of probing material at various densities.
In particular, large high-resolution dust continuum maps with a high spatial dynamic range can efficiently trace a broad range of column densities, 
while complementary spectroscopic data
are sensitive either to the densest inner regions of filaments (e.g.\ N$_2$H$^+$) or to the surrounding diffuse medium (e.g.\ CO and isotopologues).

\begin{acknowledgements}
We thank the anonymous referee for their detailed report, which helped us improving the robustness and the presentation of our results.
We are very thankful for the continuous support provided by the APEX staff
during \artemis\ operations.
FS acknowledges support from a CEA/Marie Sklodowska-Curie Enhanced Eurotalents fellowship.
DA acknowledges support by FCT/MCTES through national funds (PIDDAC) by the grant UID/FIS/04434/2019.
PP acknowledges support from FCT through the research grants UIDB/04434/2020 and UIDP/04434/2020. PP receives support from fellowship SFRH/BPD/110176/2015 funded by FCT (Portugal) and POPH/FSE (EC).
Part of this work has received support from the European Research Council under
the European Union's Seventh Framework Programme (ERC Advanced Grant Agreement
no. 291294 -- ORISTARS), from the French National Research Agency (Grant
no. ANR--11--BS56--0010 -- STARFICH), and from ``Ile de France'' regional
funding (DIM-ACAV$+$ Program). We also acknowledge support 
from the French national programs of CNRS/INSU on
stellar and ISM physics (PNPS and PCMI). 
The present study has made use of data from the Herschel Gould Belt survey (HGBS) project (http://gouldbelt-herschel.cea.fr). 
The HGBS is a Herschel Key Programme jointly carried out by 
SPIRE Specialist Astronomy Group 3 (SAG 3), scientists 
of several institutes in the PACS Consortium (CEA Saclay, INAF-IFSI Rome 
and INAF-Arcetri, KU Leuven, MPIA Heidelberg), and scientists 
of the Herschel Science Center (HSC). 
This document was prepared using the Overleaf web application, which can be found at www.overleaf.com.

\end{acknowledgements}

\bibliographystyle{aa}
\bibliography{orion_a}

\begin{appendix}

\section{Results derived with a modified dust opacity law}
\label{sec-app-alt}

\begin{figure}
   \centering
   \includegraphics[width=\hsize]{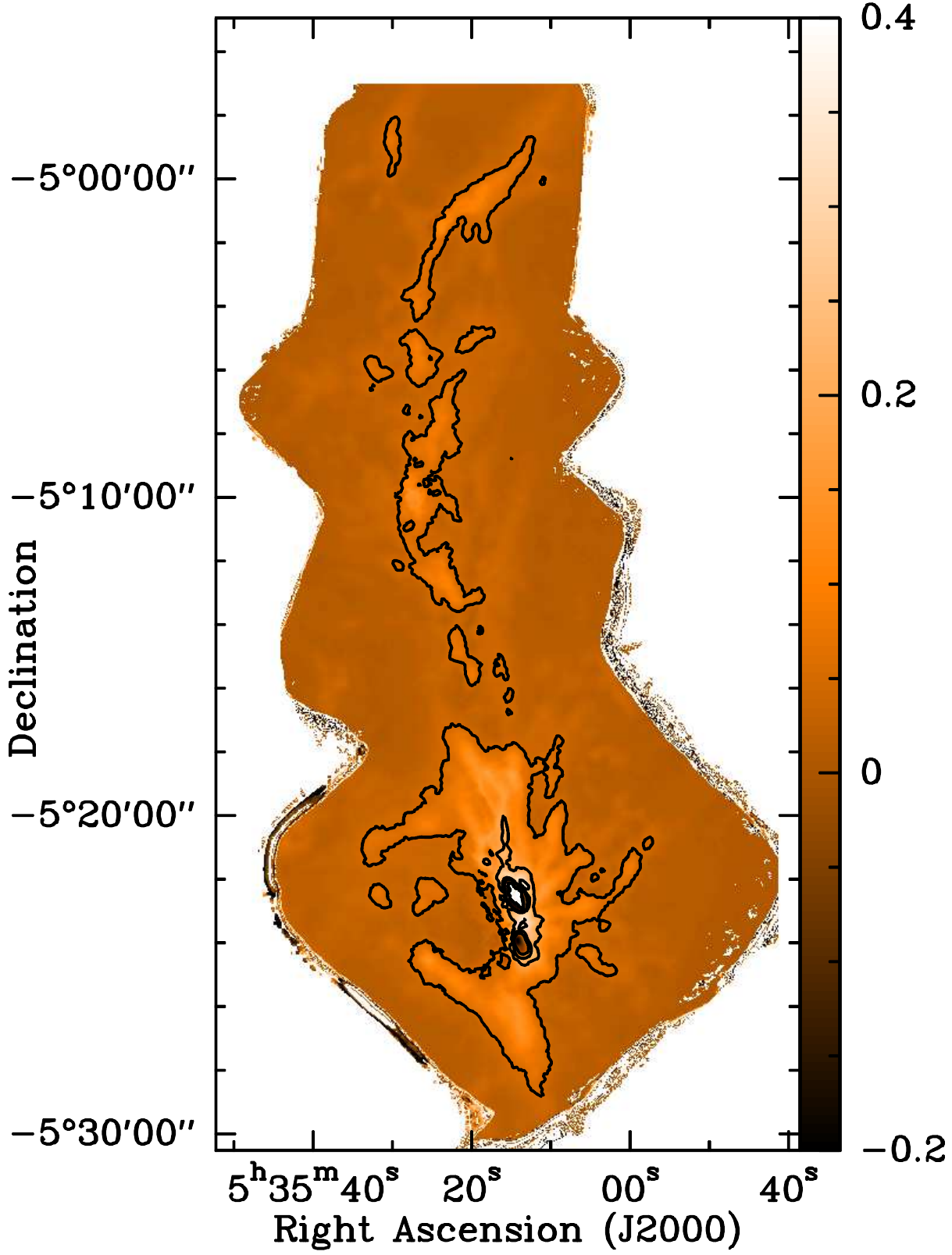}
   \caption{Map of the relative temperature difference,  $(T_d^{\rm alt} - T_d^{\rm std})/T_d^{\rm std}$, 
   between the alternative dust temperature ($T_d^{\rm alt}$), derived using the modified dust opacity
   law given by Eq.~(3),  
   and the standard dust temperature ($T_d^{\rm std}$), derived using the HGBS opacity law of Eq.~(1). 
   Contours are $+5\%$,  $+20\%$, and $+35\%$.}
   \label{fig-err-Tdust}
\end{figure}

\begin{figure}
   \centering
   \includegraphics[width=\hsize]{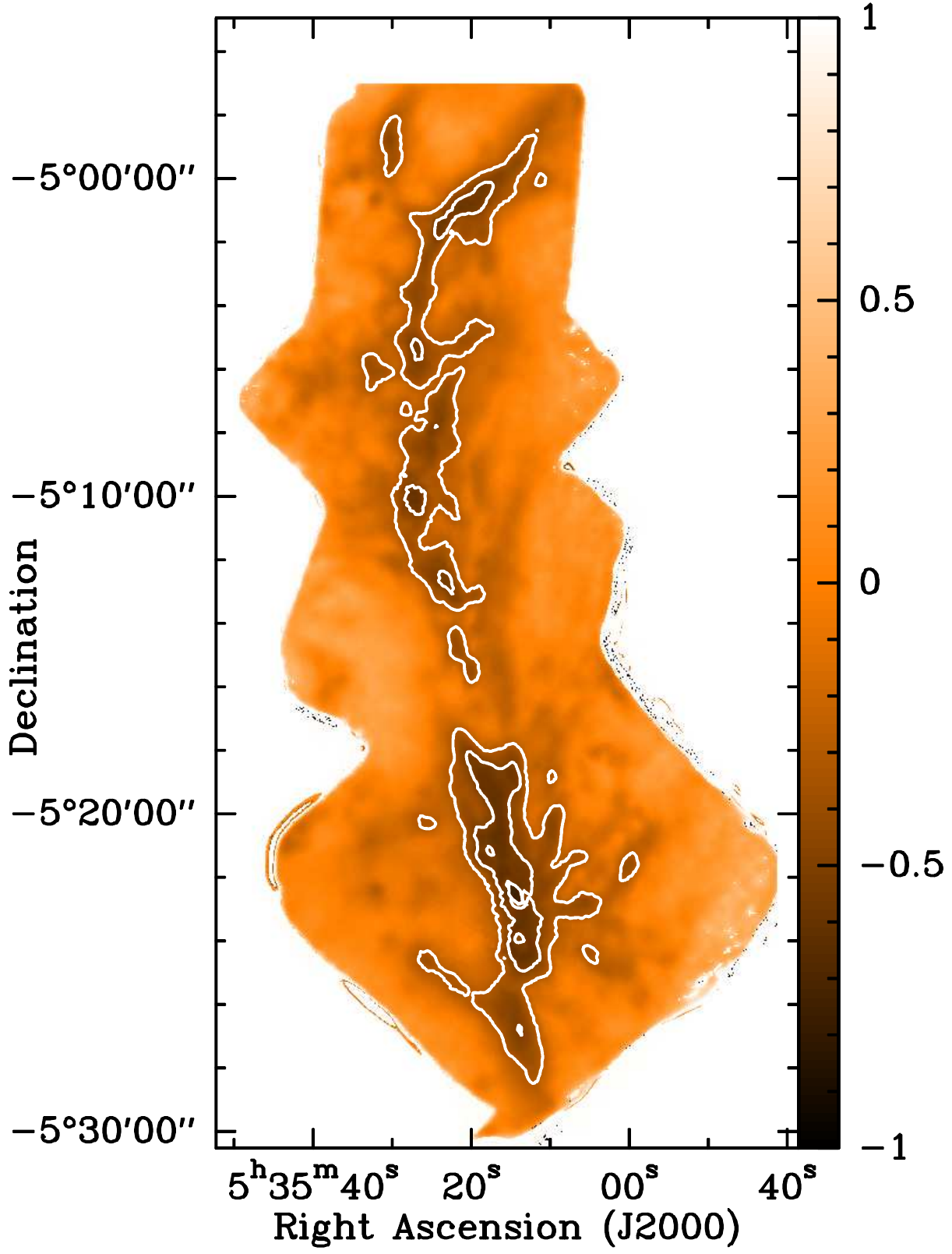}
   \caption{Map of the relative difference in column density,  $(N_{\rm H_2}^{\rm alt} - N_{\rm H_2}^{\rm std})/N_{\rm H_2}^{\rm std}$, 
   between the alternative column density ($N_{\rm H_2}^{\rm alt}$), derived using the modified dust opacity law given by Eq.~(3), 
   and the standard column density ($N_{\rm H_2}^{\rm std}$), derived using the HGBS opacity law of Eq.~(1). 
   Contours are $-30\%$,  $-50\%$, and $-70\%$. }
   \label{fig-err-coldens}
\end{figure}

In this section we present results derived from the alternative column density map,
$N_{\rm H_2}^{\rm alt}$, derived using the modified dust opacity law given by Eq.~(3)
instead of a constant value for the dust opacity as expressed in Eq.~(1).
We show maps of the relative differences in the derived dust temperature and column density
in Figs.~\ref{fig-err-Tdust} and \ref{fig-err-coldens}, respectively.
We measured the inner widths of the northern and southern parts of the main filament by
fitting Gaussian and Plummer-like profiles to the $N_{\rm H_2}^{\rm alt}$ map, using
the same method as described in Sect.~\ref{sec-method}.

\begin{figure*}
   \centering
   \includegraphics[width=\hsize]{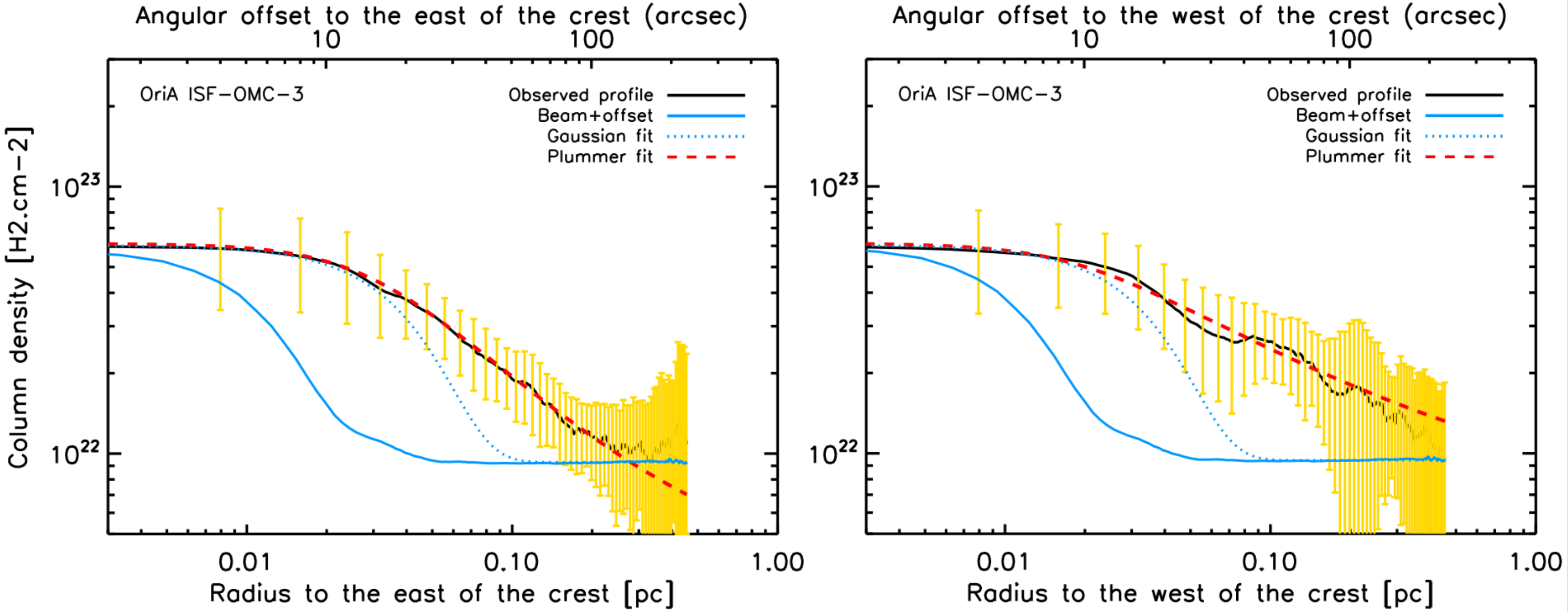}
   \caption{Median radial profiles for the northernmost part of the ISF (solid black curves) measured on the alternative column density map, on the eastern (left panel) and western (right panel) sides of the filament.
   The various lines have the same meanings as in Fig.~\ref{fig-profile_OMC3}.
  }
   \label{fig-profile_alt_OMC3}
\end{figure*}

\begin{figure*}
   \centering
   \includegraphics[width=\hsize]{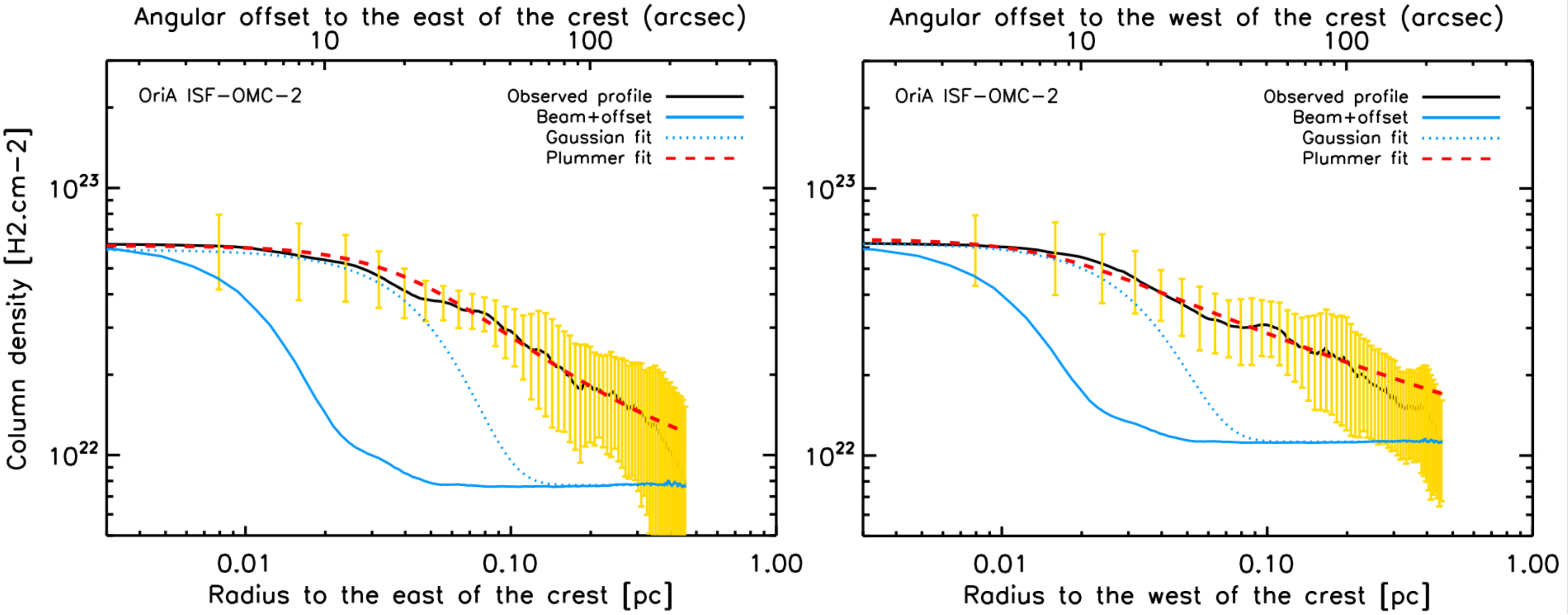}
   \caption{Same as Fig.~\ref{fig-profile_alt_OMC3} but for the portion
   of the main filament covering OMC-2.
  }
   \label{fig-profile_alt_OMC2}
\end{figure*}

\begin{figure*}
   \centering
   \includegraphics[width=\hsize]{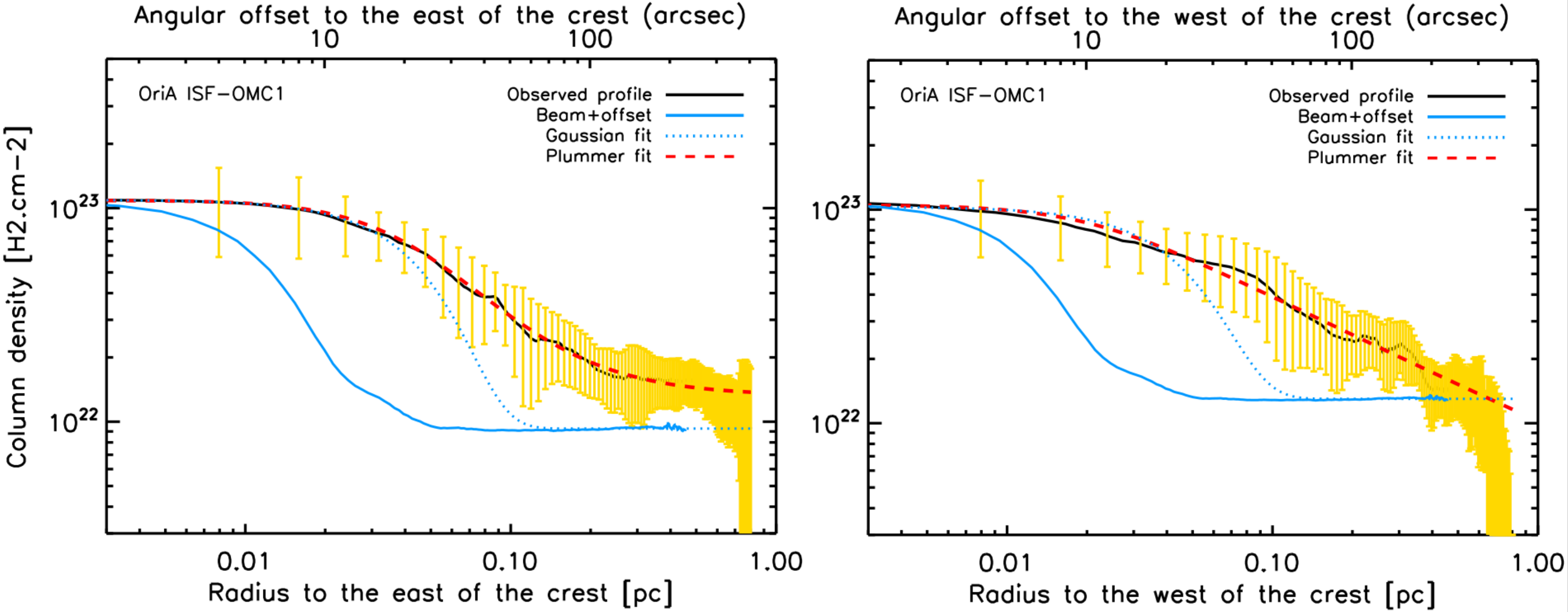}
   \caption{Same as Fig.~\ref{fig-profile_alt_OMC3} but for the southern part
   of the main filament (ISF-OMC-1).
  }
   \label{fig-profile_alt_S}
\end{figure*}

The median radial intensity profiles computed for ISF-OMC-3, ISF-OMC-2, and  ISF-OMC-1
are shown in Figs.~\ref{fig-profile_alt_OMC3}, \ref{fig-profile_alt_OMC2}, and \ref{fig-profile_alt_S}, respectively.
The parameters of the best-fit models are reported in Table~\ref{tab-alt-width},
including the results obtained when splitting these filaments into small segments that
can be associated with the fibres discussed by \citet{Hacar2018}, as detailed in
Sect.~\ref{sec-fiber} and in Table~\ref{tab-width}.

\begin{table*}
\caption[]{Median inner widths of selected filaments, as derived from both Plummer
and Gaussian fitting to the radial profiles measured in the alternative column density 
map derived assuming the modified dust opacity law of Eq.~(3).}
  \label{tab-alt-width}
  \begin{tabular}{lllll}
    \hline
    Filament & $D_{\rm flat} $ (pc)  & $p$  &  FWHM (pc) & $D_{\rm HP}^{\rm Plummer}$ (pc) \\
    \hline
    \hline
    ISF-OMC-3 & & & &  \\
    West & 0.045$\pm$0.01 & 1.6$\pm$0.3 & 0.06$\pm$0.02 & 0.09$\pm$0.01  \\
    East & 0.06$\pm$0.01  & 2.1$\pm$0.2 & 0.07$\pm$0.01 & 0.10$\pm$0.02  \\
    \hline
    ISF-OMC-2 & & & & \\
    West & 0.04$\pm$0.02 & 1.5$\pm$0.3 & 0.065$\pm$0.02 & 0.10$\pm$0.03  \\
    East & 0.075$\pm$0.02  & 2.0$\pm$0.2 & 0.08$\pm$0.02 & 0.13$\pm$0.02  \\    
    \hline
    Fibre 41 (West) & $0.05\pm0.01$ & 1.8$\pm$0.2  & 0.05$\pm$0.01 &  0.11$\pm$0.02 \\
    \hline
    Fibre 41 (East) & $0.06\pm0.02$ & 1.7$\pm$0.3  & 0.05$\pm$0.02 &  0.10$\pm$0.02 \\        
    \hline
    Fibre 44 (West) & -- &  --  & -- &  -- \\                                                                                                                                            
    \hline
    Fibre 44 (East) & ($0.07\pm0.02$) & (1.9$\pm$0.3)  & 0.06$\pm$0.03 &  0.12$\pm$0.03 \\    
    \hline
    Fibre 43 (West) & ($0.05\pm0.02$) & (1.8$\pm$0.3)  & 0.06$\pm$0.01 &  (0.09$\pm$0.03) \\                                                     
    \hline
    Fibre 43 (East) & ($0.05\pm0.02$) & (2.0$\pm$0.3)  & (0.09$\pm$0.02) &  (0.09$\pm$0.02) \\                                                   
    \hline
    Fibre 37 (West) & ($0.07\pm0.03$) & (2.0$\pm$0.5)  & (0.07$\pm$0.02) &  (0.12$\pm$0.04) \\                                           
    \hline
    Fibre 37 (East) & $0.09\pm0.03$ & 2.3$\pm$0.3  & 0.14$\pm$0.03 &  0.15$\pm$0.03 \\    
    \hline
    \hline     
    ISF-OMC-1 & & & & \\
    West & 0.04$\pm$0.01 & 1.7$\pm$0.3 & 0.08$\pm$0.02 & 0.11$\pm$0.02  \\
    East & 0.05$\pm$0.02 & 2.2$\pm$0.3 & 0.08$\pm$0.01 & 0.08$\pm$0.02  \\
    \hline
    Fibre 29 (West) & ($0.05\pm0.01$) & (2.0$\pm$0.5)  & 0.05$\pm$0.01 &  (0.08$\pm$0.02) \\ 
    \hline
    Fibre 29 (East) & ($0.07\pm0.02$) & (2.1$\pm$0.3)  & 0.07$\pm$0.01 &  0.12$\pm$0.02 \\
    \hline
    Fibre 21 (West)  & $0.04\pm0.02$ & 1.8$\pm$0.2   & 0.05$\pm$0.01 &  0.10$\pm$0.02 \\ 
    \hline
    Fibre 21 (East)  & ($0.05\pm0.02$) & (2.0$\pm$0.5)   & 0.06$\pm$0.02 &  0.07$\pm$0.01 \\         
    \hline
    Fibre 19 (West) & ($0.09\pm0.03$) & (1.6$\pm$0.4)   & 0.08$\pm$0.02 &  0.16$\pm$0.04 \\
    \hline
    Fibre 19 (East) & ($0.07\pm0.02$) & (2.0$\pm$0.3)   & 0.06$\pm$0.01 &  0.12$\pm$0.03 \\ 
    \hline
    Fibre 9 (West) & ($0.04\pm0.02$) & (2.2$\pm$0.5)   & 0.06$\pm$0.01 &   0.07$\pm$0.02 \\    
    \hline
 \end{tabular}
  \tablefoot{Values shown in parentheses denote cases with higher uncertainties in the fitted values.}
\end{table*}

\pagebreak

\vspace*{\fill} 
\newpage

\section{Influence of the filament-finding algorithm}
\label{sec-app-crests}

\begin{figure*}
   \centering
   \includegraphics[width=\hsize]{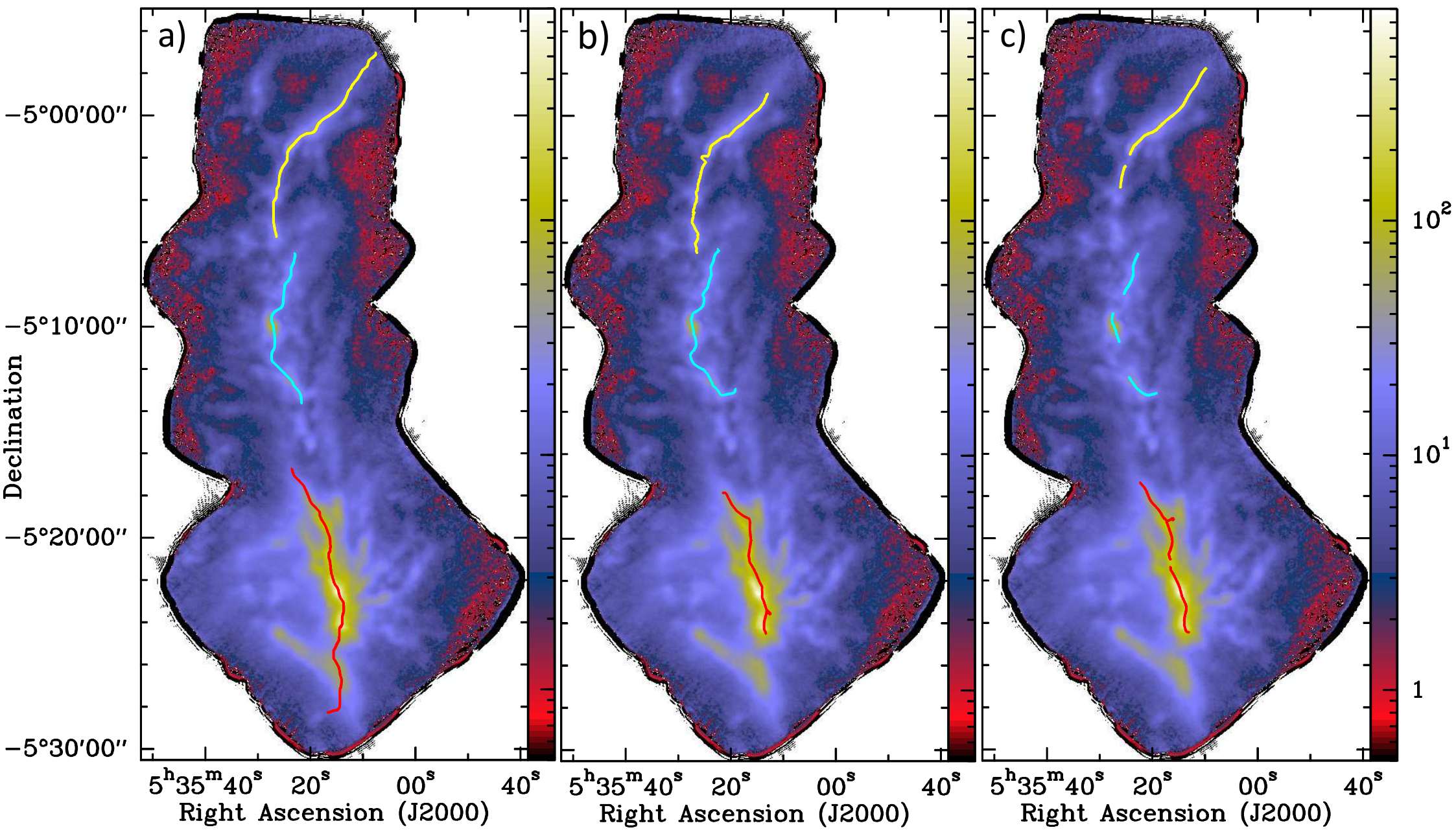}
   \caption{Comparison of the crests derived for the ISF (from top to bottom: OMC-3 portion in yellow, OMC-2 in cyan, and OMC-1 in red) 
   using three different filament-finding algorithms: DisPerSE {\bf (a)},  \textsl{FilFinder} {\bf (b)}, 
   and \textsl{getsf} {\bf (c)}. The underlying image is the same \artemis\ $+$ SPIRE 350\mic\ map as shown in Fig.~\ref{fig-skel350}.
  }
   \label{fig-skel-ISF}
\end{figure*}

To check whether the choice of the algorithm employed to trace filament crests had any significant effect on our results, 
we used two alternative algorithms, \textsl{FilFinder} \citep{Koch2015} and  \textsl{getsf} \citep{Menshchikov2021}, 
in addition to DisPerSE \citep{ref-disperse} to define the spine of the ISF. 
Figure~\ref{fig-skel-ISF} shows the results. Although \textsl{getsf} tends to break up filamentary structures into smaller pieces 
than the other two algorithms, it can be seen that the three methods agree quite well in all three portions (OMC-1, 2, and 3) 
of the ISF.

\begin{figure}
   \centering
   \includegraphics[width=\hsize]{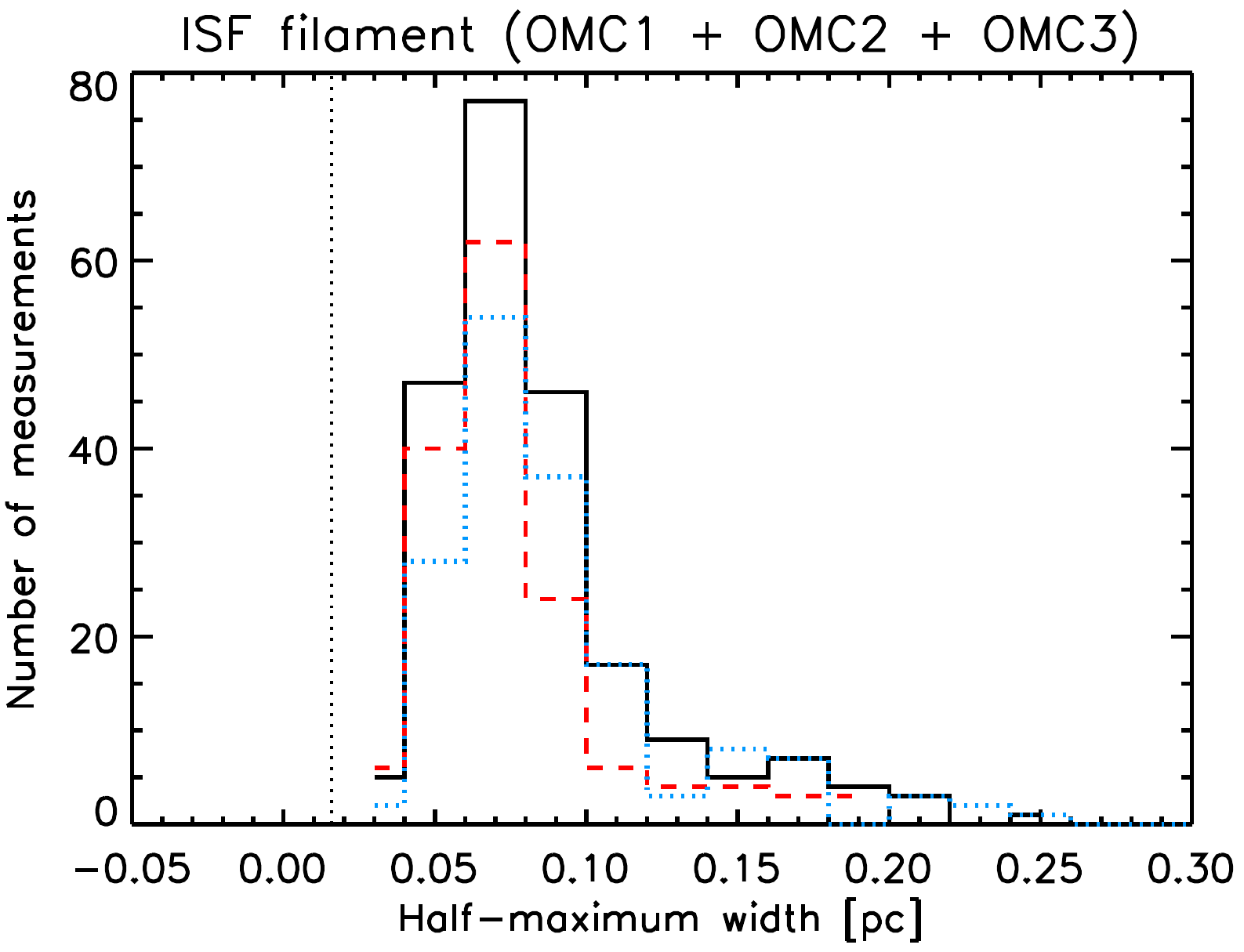}
   \caption{Comparison of the distributions of individual width measurements 
   along the whole (OMC-1 $+$ OMC-2 $+$ OMC-3) ISF obtained using 
   the crests defined by DisPerSE (same solid black histogram as in Fig.~\ref{fig-width-histo}), 
   \textsl{FilFinder}  (dotted blue histogram), and \textsl{getsf} (dashed red histogram).
  }
   \label{fig-width-getsf-filfinder}
\end{figure}

Moreover, the three distributions of individual widths measured along the whole ISF 
using the three sets of crests are very similar (see Fig.~\ref{fig-width-getsf-filfinder}). 
A Kolmogorov-Smirnov test confirms that the distribution of widths found with the DisPerSE crests (Fig.~\ref{fig-width-histo}) 
is statistically indistinguishable from the other two distributions.

\vspace*{\fill} 
\newpage

\section{Influence of temperature variations along the line of sight}
\label{sec-app-los}

\begin{figure*}
   \centering
   \includegraphics[width=\hsize]{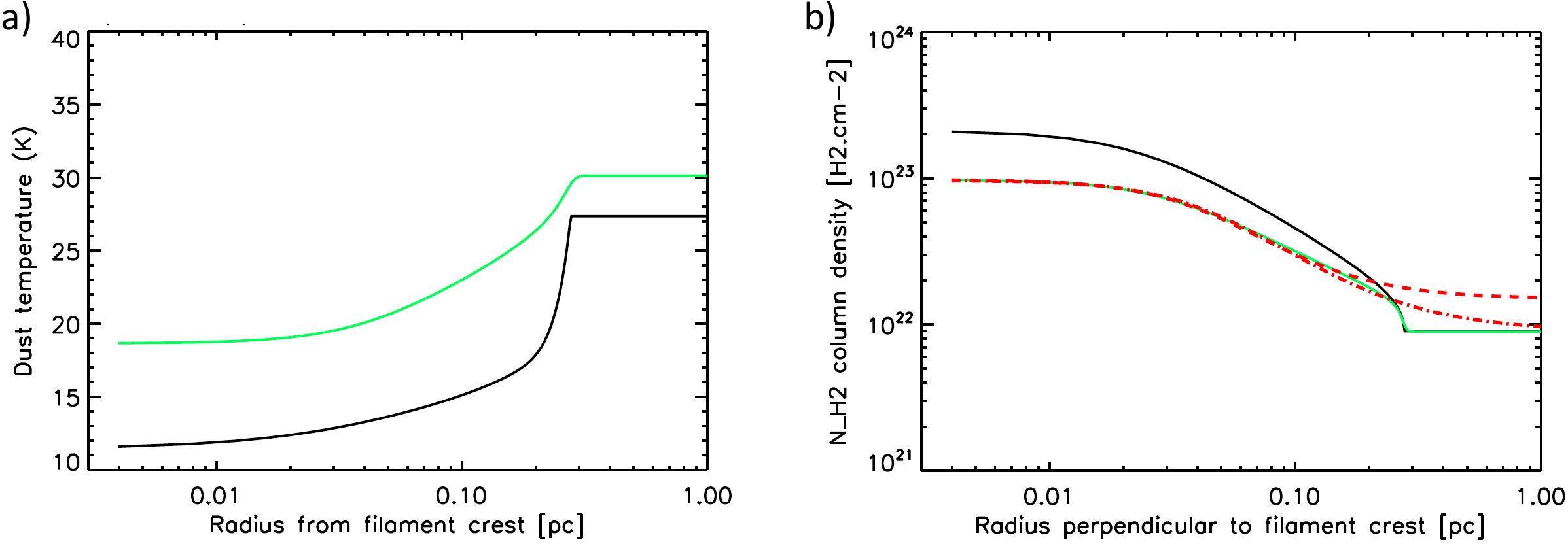}
   \caption{Synthetic dust temperature ({\bf a}) and column density ({\bf b}) profiles of a model filament with Plummer-type density structure,  
   illustrating the effect of temperature gradients along the line of sight. The true dust opacity law and the dust opacity law adopted to 
   derive temperature and column density maps from synthetic emission maps  were identical and given by Eq.~(1).
   {\bf a)} Comparison between the intrinsic model temperature profile as a function of radius (black curve) 
   and the temperature profile as a function of projected radius resulting from SED fitting and thus averaging along the line of sight (green curve). 
   {\bf b)} Comparison between the corresponding model column density profile (black curve) and 
   the column density profile resulting from SED fitting (green curve). 
   The dashed and dash-dotted red curves represent two Plummer fits to the SED column density profile, 
   which differ only from each other and from the model in the outer regions. 
  }
   \label{fig-syn-profiles}
\end{figure*}

\begin{figure*}
   \centering
   \includegraphics[width=\hsize]{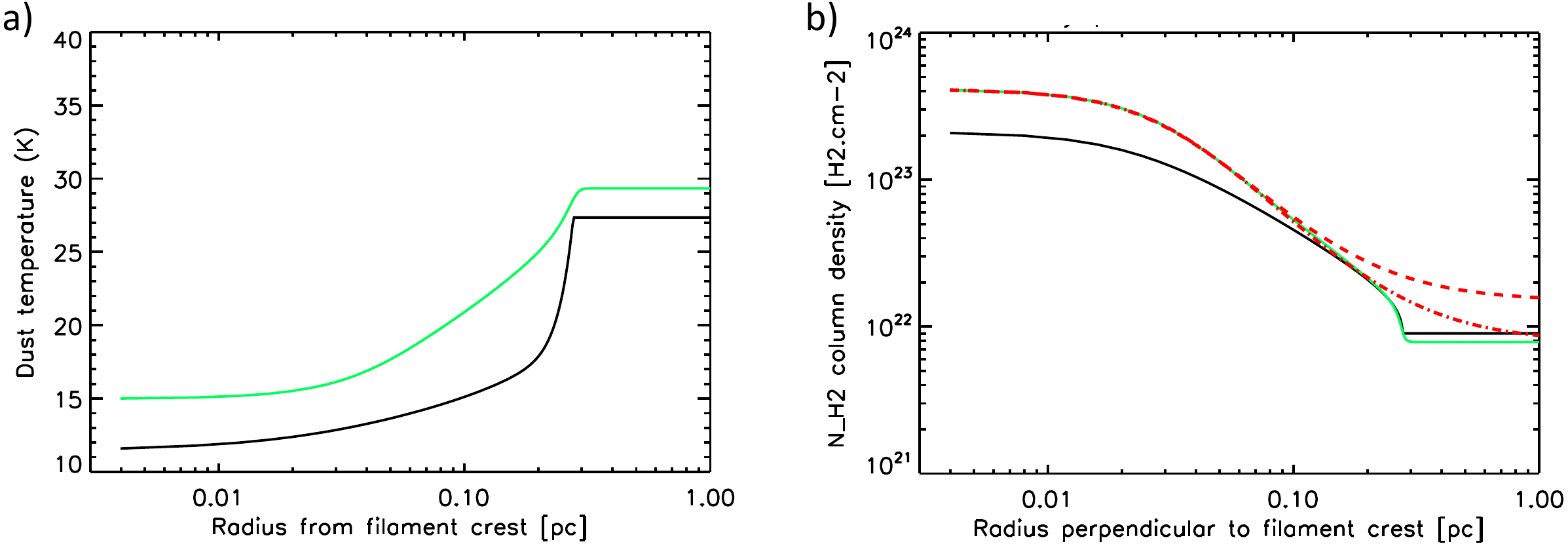}
   \caption{Same as Fig.~\ref{fig-syn-profiles}, for the same model filament 
   but assuming the true dust opacity law is given by Eq.~(3) while the dust opacity law adopted to 
   derive temperature and column density maps from SED fitting is given by Eq.~(1). 
  }
   \label{fig-syn-alt-profiles}
\end{figure*}

To evaluate the effect of temperature gradients along the line of sight 
on our results, 
we constructed synthetic images for a model filament with a Plummer-type density 
distribution (Eq.~4 with $p=2$) and a realistic  temperature distribution, assuming 
an ambient interstellar radiation field (ISRF) typical of the Orion 
region ($G_0 = 1000$)\footnote{The radiation field in the OMC-1 subregion is somewhat higher 
than the value adopted in the model described here. Accordingly, the internal dust temperature gradient 
within the OMC-1 portion of the ISF and the corresponding effects on the OMC-1 column density profiles 
may be somewhat stronger than shown in Figs.~\ref{fig-syn-profiles}~and~\ref{fig-syn-alt-profiles}.}. 
The cylindrical model filament was assumed to have a central column density of \hbox{$N_0 = 2 \times 10^{23}\ {\rm cm}^{-2}$} 
on its crest, an intrinsic half-power diameter of \hbox{$D_{\rm HP} = 0.075\,$pc}, and an outer diameter of 0.55 pc.
It was also assumed to be embedded in a background cloud of column density $N_{\rm bg} = 9 \times 10^{21}\ {\rm cm}^{-2}$ 
and, for simplicity, to lie in the plane of sky. 
The properties of this model filament are roughly similar to those derived for the ISF 
(see Sect.~\ref{sec-profile}  and Table~\ref{tab-width}).
The synthetic dust temperature profile within the model filament was calculated 
for $G_0 = 1000$ using an analytic approximation formula on a grid of
radiative transfer models performed with the MODUST code
(see \citealp{Bouwman2001} and \citealp{Andre2003}). 
Two sets  of synthetic emission maps for the model filament were generated 
at all observed {\it Herschel} and ArT\'eMiS wavelengths, assuming optically thin 
dust emission with the nominal dust opacity law given by Eq.~(1) on one hand (`B.1 test')
and the alternative dust opacity law given by Eq.~(3) on the other hand (`B.2 test'). 
Column density and dust temperature maps were then produced from 
these synthetic emission maps in the same manner as for the real data, 
by fitting a modified blackbody to the observed SEDs
on a pixel-by-pixel basis and assuming the same nominal dust opacity law 
in both cases (cf. Sect.~\ref{sec-coldens}). 
The purpose of the B.2 test is to assess the combined effect of line-of-sight temperature 
gradients and dust opacity uncertainties.


\begin{table*}
\caption[]{Results of measurement tests using synthetic maps of a Plummer model filament. }
  \label{tab-syn-width}
  \begin{tabular}{lllll}
    \hline
    Filament & $N_0$  (cm$^{-2}$) & $D_{\rm flat} $ (pc)  & $p$  & $D_{\rm HP}$ (pc) \\
    \hline
    \hline
    \hline
    Model & $2.1 \times 10^{23}$ & 0.046 & 2 & 0.075  \\
    \hline 
    B.1 fit & $0.9\times 10^{23}$ & 0.080--0.082  & 2.7--2.5     & 0.09--0.10  \\
    B.2 fit & $4.2\times 10^{23}$ & 0.027--0.028  & 2.75--2.74 & 0.06--0.062  \\
    \hline
 \end{tabular}
\end{table*}

The results of this experiment and a comparison with the input model 
are provided in Figs.~\ref{fig-syn-profiles} and \ref{fig-syn-alt-profiles}, in the form of radial temperature 
and column density profiles, and in Table~\ref{tab-syn-width}.
As expected, it can be seen in Figs.~\ref{fig-syn-profiles}a and \ref{fig-syn-alt-profiles}b 
that the temperature profiles derived from SED fitting (green curve) overestimate the intrinsic dust temperature profile 
of the model (black curve). This is due to the fact that the model filament is colder 
in its inner interior and that the temperatures derived from SED fitting represent 
line-of-sight averages that are significantly affected by the warmer outer layers of the filament. 
Accordingly, the column density profile derived assuming the correct dust opacity (green curve in Fig.~\ref{fig-syn-profiles}b) 
underestimates the intrinsic column density profile of the model (black curve in Fig.~\ref{fig-syn-profiles}b) by a factor of 2.3 
at small radii. In the B.2 case, on the other hand, the temperature effect does not quite compensate for the incorrect 
assumption about the dust opacity, and the derived column density profile (green curve in Fig.~\ref{fig-syn-alt-profiles}b)
still exceeds the intrinsic column density profile (black curve in Fig.~\ref{fig-syn-alt-profiles}b) by a factor of two
at the centre of the filament.

The shape of the derived column density profiles is nevertheless very similar 
to the shape of the model column density profile. Consequently, the half-power diameters 
derived from Plummer fits to the SED-based column density profiles (red curves in Figs.~\ref{fig-syn-profiles}b and \ref{fig-syn-alt-profiles}b) 
remain close to (within 30\% of) the true half-power diameter of the input model (see Table~\ref{tab-syn-width}). 
More precisely, the derived $D_{\rm HP}$ diameter overestimates the intrinsic half-power diameter of the model (0.075\, pc)
by only $\sim\, $30\% in test B.1 and underestimates it by only $\sim\, $20\% in test B.2. 
By comparison, the derived central column density and $p$ index values are more uncertain 
and affected by larger systematic errors (see Table~\ref{tab-syn-width}). 
These tests demonstrate that line-of-sight averaging effects have only a little impact 
on the filament width measurements reported in Sect.~\ref{sec-profile}.

\end{appendix}

\end{document}